\documentclass[journal]{IEEEtran}

\ifCLASSINFOpdf
  \usepackage[pdftex]{graphicx}
  \usepackage{epstopdf}  	
  \usepackage{soul}
  \DeclareGraphicsExtensions{.eps, .pdf, .jpeg} 
\else
 
\fi
\ifCLASSINFOpdf
\usepackage[pdftex]{graphicx}

\else

\usepackage[dvips]{graphicx}
\fi

\usepackage[table,xcdraw]{xcolor}

\usepackage[cmex10]{amsmath}
\usepackage{amsfonts}
\usepackage{amssymb}
\usepackage{algorithm}
\usepackage{algorithmic}
\usepackage{blindtext}
\usepackage{hyperref}

\usepackage{array}
\usepackage{booktabs}
\usepackage{colortbl}
\usepackage{pgfplots}
\usepackage[numbers,sort&compress]{natbib}
\usepackage{comment}
\usepackage[nolist,nohyperlinks]{acronym}

\usepackage[caption=false,font=normalsize,labelfont=sf,textfont=sf]{subfig}
\usepackage[flushleft]{threeparttable}

\begin{acronym}
\acro{5G}{Fifth Generation}
\acro{LTE}{Long Term Evolution}
\acro{AWGN}{additive white Gaussian noise}
\acro{CCDF}{complementary cumulative distribution function}
\acro{CDF}{cumulative distribution function}
\acro{CFO}{carrier frequency offset}
\acro{CP}{cyclic prefix}
\acro{CS}{cyclic shift}
\acro{CIR}{channel impulse response}
\acro{DFT}{discrete Fourier transform}
\acro{DOF}{degrees of freedom}
\acro{gNB}{Next Generation Node B}
\acro{iDFT}{inverse discrete Fourier transform}
\acro{IID}{independent identically distributed}
\acro{MGF}{moment generating function}
\acro{NR}{New Radio}
\acro{OFDM}{orthogonal frequency division modulation}
\acro{PDF}{probability density function}
\acro{PDP}{power delay profile}
\acro{PRACH}{Physical Random Access Channel}
\acro{PUSCH}{Physical Uplink Shared Access Channel}
\acro{PRB}{Physical Resource Block}
\acro{RAR}{Random Access Response}
\acro{RNTI}{radio network temporary identifier}
\acro{RA-RNTI}{random access radio network temporary identifier}
\acro{C-RNTI}{cell radio network temporary identifier}
\acro{RV}{random variable}
\acro{SCS}{subcarrier spacing}
\acro{SNR}{signal to noise ratio}
\acro{UE}{user equipment}
\acro{ZC}{Zadoff–Chu}
\acro{UL}{uplink}
\acro{RRC}{radio resource control}
\acro{RA}{random access}
\acro{NTN}{non-terrestrial networks}
\acro{PC}{power combining}
\acro{CC}{coherent combining}
\end{acronym}

\begin{document}


\title{CFO-Robust Detection for 5G PRACH under Fading Channels: Analytical Modeling and Performance Evaluation}


\author{Daniel Alarcón-Martín,
        Mari Carmen Aguayo-Torres,
        Francisco J. Martín-Vega,
        and Gerardo Gómez
\thanks{This work has been funded by MCIN/AEI/10.13039/501100011033 (Spain), by the European Fund for Regional Development (FEDER) {`\textit{A way of making Europe}´}, and the University of M\'alaga through grants PID2020-118139RB-I00 and RYC2021-034620-I.}
\thanks{The authors are with the Communications and Signal Processing Lab, Telecommunication Research Institute (TELMA), Universidad de M\'alaga, E.T.S. Ingenier\'ia de Telecomunicaci\'on, Bulevar Louis Pasteur 35, 29010 M\'alaga (Spain). (e-mail: \{dalarcon,aguayo\}@ic.uma.es)}}


\maketitle

\begin{abstract}
The \ac{PRACH} is essential for initial access and synchronization in both 5G and future 6G networks; however, its detection is highly sensitive to impairments such as high user density, large \ac{CFO}, and fast fading. Although prior studies have examined \ac{PRACH} detection, they are often restricted to specific scenarios or lack a comprehensive analytical characterization of performance.
We introduce a unified analytical framework that characterizes the statistical distribution of the received \ac{PDP} under flat Rayleigh fading and supports both \ac{CC} and \ac{PC} repetition strategies. For each strategy, we derive optimal threshold expressions and closed-form detection probabilities. Furthermore, we analyze two key cases depending on the coherence time: identical and independent channel realizations per repetition. Secondly, we exploit the correlation induced by \ac{CFO} across cyclic shifts to design a novel low-complexity detector that exploits \ac{PDP} dependencies. Numerical results indicate that \ac{PC} outperforms \ac{CC} when repetitions experience independent channels, while \ac{CC} can be preferable under identical realizations in limited settings. On the other hand, the proposed \ac{CFO}-aware detector delivers improved robustness under severe \ac{CFO} conditions.



\end{abstract}

\acresetall

\section{Introduction}
\label{sec:Introduction}
\subsection{Motivation and scope}
\IEEEPARstart{T}{he} proliferation of 5G networks marks a new era of connectivity characterized by high data rates, ultra-low latency, massive device connectivity, and enhanced reliability. A key aspect of this advancement lies in \acf{PRACH} \cite{3gpp38211}, which plays a pivotal role in establishing communication between the \acf{UE} and the network. \ac{PRACH} serves as the initial handshake in the access procedure, facilitating \ac{UL} synchronization and initial beam selection for data transmission \cite{enescu20205g}. However, due to the increasing variety of use cases, traditional \ac{PRACH} preamble detection mechanisms face significant challenges. The rising complexity of 5G scenarios introduces issues such as high user density, larger \ac{CFO}, and fast fading. To address these challenges, the 5G standard has considerably extended \ac{PRACH}, introducing a diverse set of preamble formats with increased preamble repetitions, diverse \acf{SCS} and shorter preamble length. This leads to the configuration of a larger number of roots per \ac{gNB} to reduce collision rates, increasing interference between roots, and resulting in inefficient use of radio resources. These issues are expected to persist or intensify in future 6G systems, especially when considering the integration of \ac{NTN}, spatial multiplexing, and grant-free \ac{RA}.

\subsection{Related work}

Several works in the literature address these issues by analyzing \ac{PRACH} preamble detection in scenarios such as 5G cellular networks \cite{chakrapani2020design} \cite{pitaval2020overcoming} \cite{liang2017non}. 
%
These works consider detectors that compare the \ac{PDP} of the received preambles with a unique threshold whose value is fixed and must satisfy a target false alarm probability. Peaks in the \ac{PDP} with a higher amplitude than this threshold are used to estimate the round-trip propagation delay that is used for timing advance \cite{zhu2023timing}. 
%
%
Therefore, designing a detector requires a thorough statistical analysis of the received preamble. Some of these works make simplifying assumptions about the system model, such as a single receiver antenna, a single-user transmission, or a single-root configured base station. Meanwhile, other works do not analyze the statistical performance and properties of their models.


\begin{table*}[]
\begin{threeparttable}
\centering
\caption{List of related works.}
\label{Tab:ReferencesTable}
\begin{tabular}{lllllllll}
\toprule
\multicolumn{1}{c}{Ref.}           & \multicolumn{1}{c}{Detection anal.}                  & \multicolumn{1}{c}{False alarm anal.}                 & \multicolumn{1}{c}{Channel model} & \multicolumn{1}{c}{Multi-user} & \multicolumn{1}{c}{Multi-root} & \multicolumn{1}{c}{Multi-antenna} & \multicolumn{1}{c}{Multi-repetitions} & \multicolumn{1}{c}{Year} \\ \hline
\multicolumn{1}{c}{\cite{tao2018improved}} & \multicolumn{1}{c}{\cellcolor[HTML]{67FD9A}Yes*} & \multicolumn{1}{c}{\cellcolor[HTML]{67FD9A}Yes*} & \multicolumn{1}{c}{Multipath with CFO}           & \multicolumn{1}{c}{\cellcolor[HTML]{FD6864}No}           &   \multicolumn{1}{c}{\cellcolor[HTML]{FD6864}No}         &       \multicolumn{1}{c}{\cellcolor[HTML]{FD6864}No}     &       \multicolumn{1}{c}{\cellcolor[HTML]{FD6864}No} & \multicolumn{1}{c}{2018}       \\
\multicolumn{1}{c}{\cite{wang2015multiuser}} & \multicolumn{1}{c}{\cellcolor[HTML]{FD6864}No}                            & \multicolumn{1}{c}{\cellcolor[HTML]{FD6864}No}                           & \multicolumn{1}{c}{Multipath with CFO}           & \multicolumn{1}{c}{\cellcolor[HTML]{67FD9A}Yes}           &  \multicolumn{1}{c}{\cellcolor[HTML]{FD6864}No}          &      \multicolumn{1}{c}{\cellcolor[HTML]{FD6864}No}      &      \multicolumn{1}{c}{\cellcolor[HTML]{FD6864}No}    & \multicolumn{1}{c}{2015}              \\
\multicolumn{1}{c}{\cite{yang2013enhanced}} & \multicolumn{1}{c}{\cellcolor[HTML]{67FD9A}Yes*}                            & \multicolumn{1}{c}{\cellcolor[HTML]{67FD9A}Yes*}                           & \multicolumn{1}{c}{Multipath}           & \multicolumn{1}{c}{\cellcolor[HTML]{FD6864}No}           &     \multicolumn{1}{c}{\cellcolor[HTML]{FD6864}No}       &     \multicolumn{1}{c}{\cellcolor[HTML]{67FD9A}Yes}       &       \multicolumn{1}{c}{\cellcolor[HTML]{FD6864}No}    & \multicolumn{1}{c}{2013}             \\
 \multicolumn{1}{c}{\cite{hua2013analysis}}                              &       \multicolumn{1}{c}{\cellcolor[HTML]{FD6864}No}                                          &          \multicolumn{1}{c}{\cellcolor[HTML]{FD6864}No}                                      &       \multicolumn{1}{c}{Multipath with CFO}                         &           \multicolumn{1}{c}{\cellcolor[HTML]{FD6864}No}                     &    \multicolumn{1}{c}{\cellcolor[HTML]{FD6864}No}        &       \multicolumn{1}{c}{\cellcolor[HTML]{FD6864}No}     &  \multicolumn{1}{c}{\cellcolor[HTML]{FD6864}No} & \multicolumn{1}{c}{2013}                   \\
     \multicolumn{1}{c}{\cite{ding2019analysis}}                          &     \multicolumn{1}{c}{\cellcolor[HTML]{67FD9A}Yes}                                            &          \multicolumn{1}{c}{\cellcolor[HTML]{FD6864}No}                                      &    \multicolumn{1}{c}{Flat channel}                            &         \multicolumn{1}{c}{\cellcolor[HTML]{67FD9A}Yes}                       &    \multicolumn{1}{c}{\cellcolor[HTML]{67FD9A}Yes}        &      \multicolumn{1}{c}{\cellcolor[HTML]{67FD9A}Yes}      &       \multicolumn{1}{c}{\cellcolor[HTML]{FD6864}No}  &      \multicolumn{1}{c}{2019}            \\
        \multicolumn{1}{c}{\cite{kim2017enhanced}}                       &            \multicolumn{1}{c}{\cellcolor[HTML]{67FD9A}Yes*}                                     &                   \multicolumn{1}{c}{\cellcolor[HTML]{67FD9A}Yes}                             &            \multicolumn{1}{c}{Flat channel}                    &     \multicolumn{1}{c}{\cellcolor[HTML]{67FD9A}Yes}                            &   \multicolumn{1}{c}{\cellcolor[HTML]{67FD9A}Yes}          &      \multicolumn{1}{c}{\cellcolor[HTML]{FD6864}No}      &  \multicolumn{1}{c}{\cellcolor[HTML]{FD6864}No} &        \multicolumn{1}{c}{2017}               \\
    \multicolumn{1}{c}{\cite{chakrapani2019nb}}                           &         \multicolumn{1}{c}{\cellcolor[HTML]{FD6864}No}                                        &            \multicolumn{1}{c}{\cellcolor[HTML]{FD6864}No}                                    &           \multicolumn{1}{c}{Flat channel with CFO}                     &                 \multicolumn{1}{c}{\cellcolor[HTML]{FD6864}No}               &      \multicolumn{1}{c}{\cellcolor[HTML]{FD6864}No}      &      \multicolumn{1}{c}{\cellcolor[HTML]{67FD9A}Yes}      &      \multicolumn{1}{c}{\cellcolor[HTML]{67FD9A}Yes}      & \multicolumn{1}{c}{2019}            \\
    \multicolumn{1}{c}{\cite{kim2016transmit}}                           &    \multicolumn{1}{c}{\cellcolor[HTML]{67FD9A}Yes}                                             &       \multicolumn{1}{c}{\cellcolor[HTML]{67FD9A}Yes}                                         &   \multicolumn{1}{c}{Flat channel}                             &       \multicolumn{1}{c}{\cellcolor[HTML]{67FD9A}Yes}                         &    \multicolumn{1}{c}{\cellcolor[HTML]{FD6864}No}        &      \multicolumn{1}{c}{\cellcolor[HTML]{FD6864}No}      &     \multicolumn{1}{c}{\cellcolor[HTML]{FD6864}No}      & \multicolumn{1}{c}{2016}             \\ 
    \multicolumn{1}{c}{\cite{ota2023nr}}                           &    \multicolumn{1}{c}{\cellcolor[HTML]{FD6864}No}                                             &       \multicolumn{1}{c}{\cellcolor[HTML]{FD6864}No}                                         &   \multicolumn{1}{c}{Multipath with CFO}                             &       \multicolumn{1}{c}{\cellcolor[HTML]{67FD9A}Yes}                         &    \multicolumn{1}{c}{\cellcolor[HTML]{FD6864}No}        &      \multicolumn{1}{c}{\cellcolor[HTML]{67FD9A}Yes}      &     \multicolumn{1}{c}{\cellcolor[HTML]{67FD9A}Yes}      & \multicolumn{1}{c}{2023} \\
    \multicolumn{1}{c}{Proposed}                           &     \multicolumn{1}{c}{\cellcolor[HTML]{67FD9A}Yes}                                            &         \multicolumn{1}{c}{\cellcolor[HTML]{67FD9A}Yes}                                       &      \multicolumn{1}{c}{Flat channel with CFO}                          &       \multicolumn{1}{c}{\cellcolor[HTML]{67FD9A}Yes}                         &    \multicolumn{1}{c}{\cellcolor[HTML]{67FD9A}Yes}        &     \multicolumn{1}{c}{\cellcolor[HTML]{67FD9A}Yes}       &    \multicolumn{1}{c}{\cellcolor[HTML]{67FD9A}Yes} &   \multicolumn{1}{c}{-}                  \\ \hline
\end{tabular}
\begin{tablenotes}
      \item Note: the Yes* tag means that the analysis is not done analytically.
    \end{tablenotes}
\end{threeparttable}
\end{table*}

While these assumptions are valid for LTE scenarios, they do not hold for the more robust and complex 5G scenarios. Indeed, as demonstrated in this work, the \ac{CFO} experienced by the signal must be carefully addressed to maximize preamble detection while fulfilling a given target false alarm probability. Modeling this behavior is very challenging, as it depends on the specific frequency offset of each user, the number of users transmitting in the same \ac{PRACH} occasion, and the number of roots assigned to the \ac{gNB}. 


%
The analysis of \ac{CFO} in a multi-user scenario, where the preambles are non-orthogonal, is substantially more complex than the previously analyzed cases, e.g., considering either a single-user with CFO or a multi-user case with a flat fading channel without CFO. In particular: i) in the presence of \ac{CFO}, the statistics of the \ac{PDP} depend on the user's Doppler shift; therefore, the performance depends on the type of scenario; ii) each preamble in the same \ac{PRACH} occasion interferes with the others; iii) unlike a flat channel, the interference between different roots is not constant. 

Despite its relevance, analysis of \ac{PRACH} detectors in the presence of \ac{CFO} is scarce. Most existing works avoid the need to calculate the statistics of their designed detectors, either because they consider a \ac{PRACH} configuration that is minimally affected by \ac{CFO}, or because their detection algorithms are complex. Moreover, most of the work has been done for LTE, while new aspects of 5G remain unexplored. A list of related works is given in Table \ref{Tab:ReferencesTable}. For example, in \cite{ding2019analysis}, asymptotic probability‐density‐function expressions are derived to detect Gaussian and \ac{ZC} sequences over a Rayleigh channel accounting for multiple roots, users, and antennas; although they use a non-standard \ac{ZC} length and omit any \ac{CFO}. \cite{tao2018improved} introduces a CFO‐aware framework using a generalized likelihood‐ratio test under unknown multipath, but does not analytically derive optimal detector parameters and ignores the effects of multiple receive antennas, simultaneous users, or multiple roots. \cite{hua2013analysis} formulate a transform‐domain model that incorporates user channel impulse responses, timing offsets, \ac{CFO}, and propose a multi‐step reconstruction algorithm that suppresses inter‐user interference, although they do not derive closed‐form expressions for success and false alarm probabilities. \cite{ota2023nr} investigate \ac{CC}, performed on the individual in-phase and quadrature components, and \ac{PC}, i.e. on their squared combination, in the presence of \ac{CFO} with multi-antenna and multi-users but only via simulations, resulting in a lack of a theoretical framework, which hinders practical applicability in 5G. Therefore, as seen in Table \ref{Tab:ReferencesTable}, there is a notable gap related to analysis of \ac{PRACH} detection in multi-user scenarios with \ac{CFO}, which is of paramount importance in practical settings.  




\subsection{Main contributions}
To our knowledge, theoretical analysis of success detection and false alarm probabilities in the presence of \ac{CFO} has not yet been derived for a particular detector in the case of multiple antennas, multiple users, and multiple roots despite its relevance for robust and efficient 5G cellular planning, which motivated us to address this research gap. The main contributions of this work are:

\begin{enumerate}
   \item We present a mathematical framework to obtain the \ac{PDP} statistics, considering the Rayleigh channel case and its extension including \ac{CFO} for a multi-antenna, multi-user, multi-root scenario.
   \item We analyze different types of signal combining for the preamble reception, showing results for different \ac{PRACH} formats.
   \item We derive closed-form expressions for success detection and false alarm probabilities of a novel statistically-informed detector that takes \ac{CFO} into account.
   \item Finally, we discuss and illustrate the usefulness of the expressions by analyzing success and false alarm probabilities versus those of a conventional detector. 
\end{enumerate}

The designed detector also aims to improve the inefficient use of radio resources. As mentioned above, due to \ac{CFO} and the need for more roots, interference between them increases, and consequently, false alarms and the radio resources allocated to respond to them also increase. This effect can be mitigated by using formats with larger \ac{SCS}, but then more bandwidth will be used for \ac{PRACH} transmission, and the effects of multiple roots will still persist. Our \ac{CFO}-based detector aims at reducing these effects avoiding the need to increase \ac{SCS}  and thus improving network efficiency.

The rest of the work is structured as follows. In Section \ref{sec:PRACH_Overview}, an overview of \ac{PRACH} is provided, explaining its parameters and its role within the \ac{RA} procedure. In Section \ref{sec:SystemModel}, the system model is described, detailing the transmission and reception of \ac{PRACH}. In Section \ref{sec:transitionProb}, the derivation and analysis of \ac{PDP} statistics for different signal combiners are performed. The optimal threshold for the case of a flat channel without \ac{CFO} is presented in Section \ref{sec:OptimalDetectorWithoutCFO}. In Section \ref{sec:DetectorDesignCFO}, a novel detector developed in the presence of \ac{CFO} is explained. Numerical results are presented in Section \ref{sec:NumericalSimResults}. Finally,  conclusions are drawn in Section \ref{sec:Conclusions}.

\section{PRACH overview} 
\label{sec:PRACH_Overview}
In 5G \ac{NR}, the \ac{UE} initiate communication with the network by transmitting a randomly selected preamble during specific time-frequency resources known as \ac{PRACH} occasions. These occasions are defined and broadcasted by the \ac{gNB} and are crucial to coordinating contention-based access in a scalable and efficient manner. Each \ac{PRACH} occasion is characterized by several configuration parameters \cite{3gpp38211}:
\begin{itemize}
    \item The \ac{PRACH} \textit{ConfigurationIndex} parameter determines the time and frequency resources allocated within a radio frame.
    \item  The \ac{SCS}, defining the numerology for \ac{PRACH} transmission, supports a wide range such as 1.25 kHz, 5 kHz or from 15 kHz to 240 kHz, to accommodate diverse deployment scenarios.
    \item The \ac{PRACH} \textit{Format} dictates the sequence length, \ac{CP}, sequence repetitions, and time-domain guards, enabling flexible adaptation to requirements such as coverage or latency.
    \item The \ac{PRACH} \textit{SequenceIndex} parameter specifies which root sequence is used as the basis for all preambles.
\end{itemize}

\ac{PRACH} preambles are based on \acf{ZC} sequences, which exhibit excellent autocorrelation and low cross-correlation properties. The network determines a root and applies \acp{CS} to generate multiple preambles while preserving the zero-correlation zone. If additional preambles are required, multiple roots can be configured, allowing expansion in the code-domain. Combined with multiplexing in the time and frequency domains through separate \ac{PRACH} occasions, this approach allows massive user access with minimal contention.

Power control is performed based on the estimated path-loss and the target receive power parameter $P_0$ configured by the cell \cite{38.213}. Then, the \ac{gNB} detects preambles by computing cross-correlation and derives a \ac{RA-RNTI} that identifies the \ac{PRACH} occasion. It transmits the \ac{RAR}, which includes a timing advance command, an \ac{UL} grant for further communication, and possibly initiates contention resolution. The \ac{UE} monitors for a \ac{RAR} from the \ac{gNB} to respond with an \ac{UL} transmission named \textit{message} $3$ (Msg3). 

In cases of preamble collision, where multiple \acp{UE} select the same resources, the \ac{gNB} responds with a common \ac{RAR}. Only one \ac{UE}'s subsequent message may be successfully decoded, and contention is resolved through a dedicated message (Msg4) carrying the \ac{C-RNTI} of the successful \ac{UE}. While undetected transmissions result in access retries, false alarms waste resources; thus, 5G \ac{NR} specifications strictly regulate detection and false alarm probabilities to maintain access efficiency.

\subsection{Notation} 
\label{subsec:Notation}

We consider sequences of finite length $\ell$,  $\{x[n]\}_{n=0:\ell-1}$. In vector form, ${\mathbf{x}} = \left\{ {x[n] };n = 0,1,...,\ell  - 1 \right\}$. 
\Ac{DFT} of the finite length discrete sequence is ${\mathbf{X}} = \text{DFT}\{{\mathbf{x}}\}  = \{ X[\nu {\text{] }};\nu  = 0,1,...,\ell  - 1 \}$ with $X[\nu ] = \sum\limits_{n = 0}^{\ell - 1} x [n]e^{-j\frac{{2\pi n\nu }}{{\ell}}}$. The periodic cross-correlation between two finite-length discrete sequences, ${\mathbf{x}}$ and ${\mathbf{y}}$ of the same length can be expressed as ${\mathbf{c}} = {\mathbf{x}} \star {\mathbf{y}}$ and its term evaluated as: $c [k] = \frac{1}{\ell} \sum_{n=0}^{\ell-1} x[n] y^{*}[(n-k)_{\ell}],$
where $\star$ means the circular cross-correlation operation, $(\cdot)^{*}$ denotes the complex conjugate, and \(\left( \cdot \right)_{\ell}\) denotes the modulus-\(\ell\) operation. It can also be evaluated in the frequency domain using the \ac{iDFT} as ${\mathbf{c}} = {\mathbf{x}} \star {\mathbf{y}} = {\text{iDFT}}\{ {\mathbf{X}}\odot{\mathbf{Y}}^{*}\}$, with $\odot$ meaning point-to-point multiplication.  

Notation relative to random variables and vectors, as well as some distributions useful along this text, can be found in the Appendix. 

\subsection{Zadoff-Chu sequences}
\label{sec:Zadoff-Chou}
\ac{PRACH} preambles in 5G are based on \ac{ZC} sequences, a class of polyphase sequences defined as \cite{3gpp38211}:
\begin{equation}
\label{eq:ZCsequeceRoot}
    \mathbf{x}_u  = \{ x_{u} [n] = e^{-\frac{ j\pi u n(n + 1)}
    {\ell} } ; n = 0,1,...,\ell  - 1 \},
\end{equation}
where \(\ell\) is the sequence length and $u$ is the root of the sequence, relatively prime to $\ell$.

A set of random access sequences with the same root can be obtained by cyclically shifting the original \ac{ZC} sequence for a given $u$ root, and \ac{CS}, $\kappa_{\mathrm{v}}$, as follows \cite{3gpp38211}:
\begin{equation}
\label{eq:ZCsequencewcs}
    {\mathbf{x}}_{u,\mathrm{v}}  = \left\{{x_{u,\mathrm{v}} [n] \triangleq x_u [(n + \kappa _\mathrm{v} )_{\ell } ];n = 0,1,...,\ell  - 1} \right\},
\end{equation}
where \ac{CS} $\kappa_{\mathrm{v}}  \in \mathcal{K}$ is one of the \acp{CS} available in the set that contains the permitted \acp{CS} relative to the original \ac{ZC} sequence in eq. (\ref{eq:ZCsequeceRoot}) of root $u$. The set of valid $u$-roots is given in [\cite{38.211} Tables 6.3.3.1-3 to 6.3.3.1-4B], and are chosen so that the interference between them is minimal. 

\subsection{\ac{ZC} sequences in 5G \ac{NR}}
\label{subsec:cross1}
In principle there are $\ell-1$ different cyclic shifts. However, a limited set $\mathcal{K}$ is employed in 5G \cite{3gpp38211}, defined with a certain \ac{CS} granularity, $n^{(\mathrm{CS})}$: 
\begin{equation}
\label{eq:CS}
    \mathcal{K} = \left\{ {\kappa_{\mathrm{v}}  = \mathrm{v}  n^{(\mathrm{CS})}; \mathrm{v}  = 0,1,...,\left\lfloor {\frac{\ell}{n^{(\mathrm{CS})}}} \right\rfloor  - 1} \right\}.
\end{equation}
Consequently, there are $\lfloor {\frac{\ell}{n^{(\mathrm{CS})}}} \rfloor$ different \acp{CS} per root. Thus, if a sequence is delayed, a guard time is kept around those employed \acp{CS} at the expense of limiting the number of different sequences for a certain root. This guard is named zero-correlation zone, its width is $n^{(\mathrm{CS})}$, and it is determined by a \ac{RRC} message containing the  \textit{zeroCorrelationZoneConfig} parameter \cite{3gpp38211}. It depends on the range of the area to be served, as it tries to mitigate imbalances such as timing offsets and channel delay spread. In contrast, this reduces the number of preambles per root.

\subsection{Cross-correlation between \ac{ZC} sequences}
\label{subsec:cross2}
We define the cross-correlation between two cyclic shifted \ac{ZC} sequences as:
\begin{equation}
\label{eq:corr}
{\mathbf{c}}_{u_1 ,\mathrm{v}_1 ,u_2 ,\mathrm{v}_2 }^{}  = {\mathbf{x}}_{u_1 ,\mathrm{v}_1 }  \star {\mathbf{x}}_{u_2 ,\mathrm{v}_2 }. 
\end{equation}
For a given root $u_0$, the \ac{ZC} sequence possesses an ideal periodic autocorrelation property: the sequences obtained from cyclically shifting the same \ac{ZC} sequence are orthogonal, i.e., they exhibit zero cross-correlation and perfect autocorrelation \cite[Ch. 3]{enescu20205g}: 
\begin{equation}
\label{eq:corrSameRoot}
c_{u_0,\mathrm{v}_1 ,{u_0},\mathrm{v}_2 }^{} [k] = \delta [(k - (\ell + \kappa _{\mathrm{v}_{2}}  - \kappa_{\mathrm{v}_{1}}))_\ell ],k = 0, ...,\ell  - 1,
\end{equation}
where $\delta[n]$ is the Kronecker delta, which is $1$ for $n = 0$ and $0$ otherwise.

For different root values, the cross-correlation is not zero, but is instead equal to a constant value \cite[Ch. 3]{enescu20205g}:
\begin{equation}
\label{eq:corrDiffRoot}
 c_{u_1 ,\mathrm{v}_1 ,u_2 ,\mathrm{v}_2 } [k] =  \frac{1}
{{\sqrt {\ell } }},k = 0, ..., \ell  - 1,u_1  \ne u_2. 
\end{equation}


\subsection{Cross-correlation between frequency-shifted \ac{ZC} sequences}
\label{subsec:cross3}
A shift in frequency $\epsilon$ produces a modulation, i.e., it results in a new sequence given by: 
\begin{equation}
\label{eq:shiftSequence}
{\mathbf{x}}_{u,\mathrm{v}}^{\epsilon}  = \left\{ {x_{u,\mathrm{v}} [n]  e^{\frac{{j2\pi \epsilon n}} {{\ell }}};n = 0,1,...,\ell  - 1} \right\}.
\end{equation}
We define the cross-correlation between one shifted sequence and another reference sequence as: 
\begin{equation}
\label{eq:corrshiftSequence}
{\mathbf{c}}_{u_1, \mathrm{v}_1, u_2, \mathrm{v}_2}^{\epsilon_{1}}={\mathbf{x}}_{u_1,\mathrm{v}_1}^{\epsilon_1}  \star {\mathbf{x}}_{u_2,\mathrm{v}_2}.
\end{equation}
Then, the cross-correlation between the sequence ${\mathbf{x}}_{u_g,\mathrm{v}_g}^{\epsilon_{g}}$ and a non-shifted reference sequence ${\mathbf{x}}_{u_0, 0}$ can be evaluated as: 
\begin{multline}
\label{eq:corrShiftedSameRoot}
{\mathbf{c}}_{u_g, \mathrm{v}_g, u_0, 0 }^{\epsilon_{g}} [k]= \frac{1}
{\ell} \sum_{n=0}^{\ell-1} {x_{u_g ,\mathrm{v}_g } [n]  e^{\frac{{j2\pi n \epsilon_{g} }}{{\ell }}}  x_{u_0 ,0}^* [(n - k)_{\ell } ]}=\\
=\frac{1}{\ell} e^{\frac{-\pi u_{g} \mathrm{v}_{g} (\mathrm{v}_{g}+1) + \pi u_{0} k (k-1)}{\ell}}  \mathrm{G}(u_{0}-u_{g}, \alpha(k)-\epsilon_{g}),
\end{multline}
where $k=0,...,\ell-1$; $ \alpha(k) = u_{g} \mathrm{v}_{g} + u_{0} k + \frac{1}{2} (u_{g}+u_{0})$; and $\mathrm{G}(a,b)$ is a quadratic Gauss sum \cite{berndt1998gauss}, defined as:
\begin{equation}
\label{eq:quadraticGaussSum}
     \mathrm{G}(a, b) = \sum_{n=0}^{\ell-1} e^{\frac{j\pi an^{2}}{\ell}-\frac{j2\pi bn}{\ell}}.
\end{equation}

\section{System model}
\label{sec:SystemModel}

We consider a single-cell network in which the \ac{gNB} configures \(n^{(\mathrm{root})}\) root sequences, collected in the set \(\mathcal{R}\), to generate \ac{PRACH} preambles. In a given \ac{PRACH} occasion, \(n^{(\mathrm{dev})}\) devices attempt \ac{RA} for \ac{UL}  transmissions. Let \(n^{(\mathrm{dev})}_{u_{p}}\) denote the number of devices selecting root index  \(u_{p} \in \mathcal{R}\), for \(p = 0, 1 ..., n^{(\mathrm{root})}-1\), to generate their preambles. We denote ${\cal D}$ the set of all active devices transmitting at the considered \ac{PRACH} occasion.  The total number of devices is thus \(n^{(\mathrm{dev})} = \sum_{p=0}^{n^{(\mathrm{root})}-1} {n^{(\mathrm{dev})}_{u_{p}}}\). Without loss of generality, we consider $u_0$ as the reference root, and $\bar{\mathcal{D}}_{u_{0}}$ is the subset of $\mathcal{D}$ formed by $n^{(\mathrm{inter})}_{u_0}=n^{(\mathrm{dev})}-n^{(\mathrm{dev})}_{u_0}$ active devices that are not transmitting with $u_{0}$, but with a different root. 

\subsection{\ac{PRACH} transmitter}
\label{subsec:txor}
\begin{figure}[tb!]
\centering
\includegraphics[width=\columnwidth,trim= 
      0pt  
      0pt   
      0pt  
      0pt   
    ,
    clip]{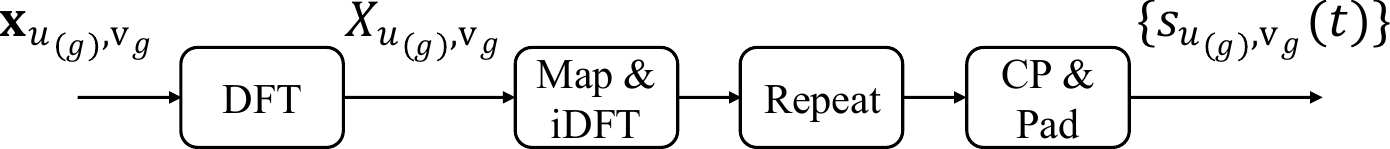}
\caption{The $g$-th device \ac{PRACH} transmitter.}
\label{fig:TxorMC}
\end{figure}

A \ac{PRACH} transmitter follows the next steps (see Fig. \ref{fig:TxorMC}):
\begin{enumerate}
    \item The $g$-th device randomly selects a cyclic shift from one of the configured roots $u_{(g)}$. Then, it generates a \ac{ZC} sequence, applying the selected \ac{CS} $\kappa_{\mathrm{v}_g}$ to derive the actual sequence $\mathbf{x}_{u_{(g)},\mathrm{v}_g}$.
    \item Obtain its frequency domain representation ${X}_{u_{(g)},\mathrm{v}_g}$ as its \ac{DFT}.
    \item Allocate the sequence at the $\ell$ subcarriers appropriate for the corresponding \ac{PRACH} occasion.
    \item Perform single carrier modulation using an inverse \acs{DFT}. Transmit the result $n^{(\mathrm{rep})} \geq 1$ times, as described by the selected format. A \ac{CP} is appended after the last repetition; intermediate \acp{CP} are not required, as the previous repetition provides the required guard interval.
    \item Apply the open-loop power control \cite{38.213} by adjusting the transmitted power such that the Preamble Received Target Power is achieved at the gNB.
\end{enumerate}

The transmitted sequence in the time domain $s_{u_{(g)} ,{\mathrm{v}}_g } (t)$ can be written as \cite{3gpp38211}: 
\begin{multline}
\label{eq:s_un_vn}
s_{u_{(g)} ,{\text{v}}_g } (t) = \\ \beta _g^{{\mathrm{(RA)}}} \sum\limits_{\nu  = 0}^{\ell - 1} {X_{u_{(g)} ,\text{v}_g } } [\nu ] e^{j2\pi (\nu + K k_{1} + \bar{k} ) \Delta f^{{\mathrm{(RA)}}} (t - t^{{\mathrm{(CP)}}} - t_\mathrm{start})}, 
\end{multline}
with
\begin{equation*}
\label{eq:Khat}
    K = \Delta f^{{\mathrm{(PUSCH)}}} / \Delta f^{{\mathrm{(RA)}}},
\end{equation*}
\begin{equation*}
\label{eq:tstart}
    t_\mathrm{start} 	\leq t < t_\mathrm{start} + (N_{u}+N_\mathrm{CP})T_\mathrm{c},
\end{equation*}
where $\Delta f^{{\mathrm{(RA)}}}$ and $\Delta f^{{\mathrm{(PUSCH)}}}$ are the \ac{SCS} of the \ac{PRACH} and \ac{PUSCH} respectively; $k_{1}$ is the starting subcarrier index, $\bar{k}$ is a default value that depends on the format and controls the repetitions of the sequence, $t_\mathrm{start}$ adjusts the start of the preamble to its starting symbol, $N_{u}$ is the useful number of samples, $N_\mathrm{CP}$ is the \ac{CP} number of samples, $T_\mathrm{c}$ is the time unit for 5G \ac{NR} (sampling period), $t^{{\mathrm{(CP)}}}$ is the \ac{CP} duration computed as $N_\mathrm{CP}T_\mathrm{c}$, and $\beta _g^{{\mathrm{(RA)}}}$ is the amplitude scaling factor in order to conform to the transmit power \cite{38.213}. We assume perfect open-loop power control and time-alignment.

\subsection{Channel model}
\label{subsec:channel}
Each device transmits the preamble signal over an independent channel to the $n^{(\mathrm{ant})}$ receiving antennas at the \ac{gNB}. We assume a flat-fading channel, implying that channel gains remain constant across the \({\ell}\) allocated subcarriers, that is, the coherence bandwidth encompasses all the subcarriers used to transmit the sequence. For the $g$-th device, the channel toward the $i$-th antenna during the $m$-th repetition is denoted by $h_g^{m,i}$. The vector containing the channel realizations for all active users at the considered \ac{PRACH} occasion is given by ${\bf{h}}$.
We assume channel realizations to be independent across devices and antennas, that is, $h_g^{m,i} \perp\!\!\!\perp h_g^{m,i'} \perp\!\!\!\perp h_{g'}^{m,i}$. However, we consider two key cases depending on the coherence time across repetitions:
\begin{itemize}
    \item Independent channel realizations at different repetitions $h_g^{m,i}  \perp\!\!\!\perp h_g^{m',i}$. 
    \item  Identical channel realizations at different repetitions, that is, the coherence time is longer than the duration of the whole \ac{PRACH} sequence and repetitions $h_g^{m,i} = h_g^{m',i}$.
\end{itemize}

Moreover, we model the channel's possible Doppler shift as a \ac{CFO} between the transmitter and the receiver, which shifts the signal in frequency. Each device suffers a different \ac{CFO}, $\epsilon_g  = \frac{{\Delta f_g }}{{\Delta f^{(\mathrm{RA})} }}$, where $\epsilon_{g}$ is the normalized \ac{CFO} and $\Delta f_{g}$ is the absolute frequency offset of device $g$. We name the set of all \acp{CFO} as $\varepsilon= \{\epsilon_1 ,...,\epsilon_{n^{(\mathrm{dev})} } \}$.

\begin{figure*}[t!]
\centering
\includegraphics[width=\textwidth,trim= 
      2pt  
      2pt   
      10pt  
      5pt   
    ,
    clip]{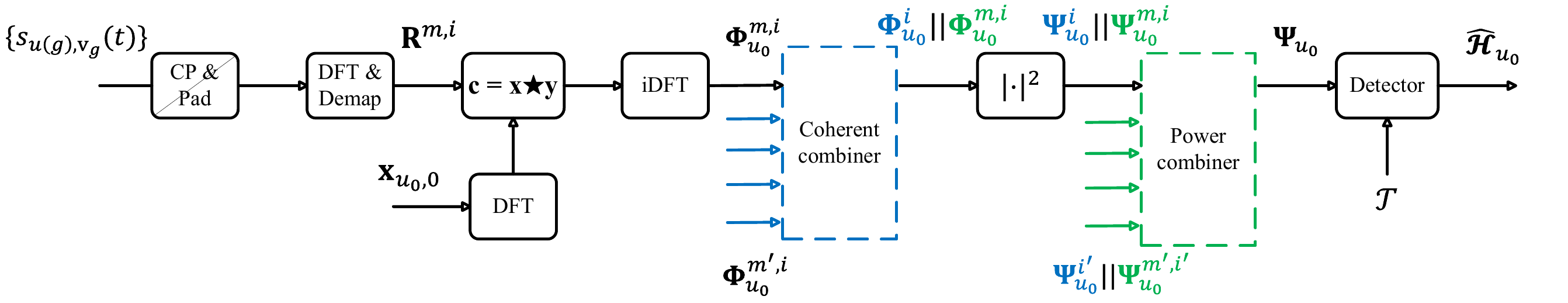}
\caption{Full frequency domain representation of the PRACH receiver.}
\label{fig:RxorMC}
\end{figure*}

Under ideal power control assumption, all links can be modeled as complex Gaussian channels with equal power, $h_g^{m,i} \sim {\cal CN}(0,\sigma _h^2 )$. For convenience, we normalize the channel to $\sigma _h^2  = 1$ without loss of generality. The noise at each receiving antenna is modeled as $w^{m,i} \sim {\cal CN}(0,\sigma_w^2 )$. Accordingly, the per-antenna \ac{SNR} is $\frac{1}{{2 \sigma _w^2 }}$. We also assume that all devices are sufficiently synchronized to remain within the receiving window, and that any residual delay can be absorbed as a phase shift in the channel response. 

\subsection{\ac{PRACH} receiver}
\label{subsec:rxor}

Fig. \ref{fig:RxorMC} illustrates the representation of the \ac{PRACH} processing for $i$-th antenna, $m$-th repetition, and root $u_0$ under test. First, both the \ac{CP} and padding are removed from the received signal. Then, the subcarriers corresponding to the \ac{ZC} sequence are demapped via the DFT. Next, cross-correlation is performed between the received sequence and the reference sequence associated with the root configured at the \ac{gNB}. If there is more than one antenna or sequence repetition, a signal combination is applied that may be coherent or power-based. Finally, a threshold-based detector determines whether a signal has been transmitted.

To facilitate analysis and without loss of generality, we assume $k_{1}=0$ in eq. \eqref{eq:s_un_vn}, which implies that the preamble is transmitted in the first set of \acp{PRB}. Furthermore, since we assume that the channel gains of $\ell$ subcarriers are equal, as we consider a flat fading channel, the aggregated received signal from all users can be written in the frequency domain as: 
\begin{equation}
\label{eq:ReceivedSignalFrequency}
    R^{m,i} [\nu] = \sum\limits_{g = 1}^{n^{(\mathrm{dev})} } {H_g^{m,i} X_{u_{(g)} ,\mathrm{v}_g} [\nu]}  + W^{m,i} [\nu].
\end{equation}
where $W^{m,i} [\nu]$ is the additive noise term in frequency; $X_{u_{(g)} ,\mathrm{v}_g} [\nu]$ is the $\nu$-th element of the sequence transmitted by the $g$-th user in the frequency domain; and $H_g^{m,i}$ denotes the channel frequency response from the $g$-th user to the $i$-th antenna during repetition $m$.

The correlation at each antenna $i$ and repetition $m$ between the received signal and the reference sequence for the evaluated root ($u_0$ without loss of generality) is denoted as $\Phi _{u_0}^{m,i} [k]$. Instead of evaluating different correlations for all \ac{CS} sequences, it is evaluated in the frequency domain with the reference sequence at zero delay, $X_{u_0,0}$, and the whole vector of correlations is kept after the \ac{iDFT}. 


Then, taking into account the linearity of \ac{iDFT} as well as the circular correlation, $\Phi_{u_0}^{m,i} [k]$ can be written as \cite{kim2017enhanced}: 
\begin{equation}
\label{eq:phi_t,i}
\Phi_{u_0}^{m,i} [k] = \mu _{u_0 }^{m,i} [k]({\bf{h}}) + z^{m,i} [k]; k = 0,...,\ell-1 
\end{equation}
being 
\begin{equation}
\label{eq:mu_t,i}
\mu _{u_0 }^{m,i} [k]({\bf{h}}) = \sum\limits_{g \in {\cal D}}^{} {h_g^{m,i}  c_{u_{(g)} ,\mathrm{v}_g ,u_0 ,0}^{\epsilon_g } [k]},     
\end{equation}
with $z^{m,i} [k] = \frac{1}{{\ell}}\sum\limits_{n = 0}^{\ell- 1} {w^{m,i} [n] x_{u_0 ,0}^* [(n - k)_{\ell } ]}$ complex Gaussian noise, $z^{m,i}[k] \sim {\cal{CN}}(0,\sigma_w^2/\ell)$, and $c_{u_{(g)} ,\mathrm{v}_g ,u_0 ,0}^{\epsilon_g } [k]$ the correlation of the reference signal with that transmitted by the $g$-th device and shifted in frequency $\epsilon_g$ as given in eq. \eqref{eq:corrShiftedSameRoot}.


\subsection{Decision statistic}
\label{subsec:dimensionality}



The different behavior in $\Phi_{u_0}^{m,i} [k]$ enables distinguishing whether there exists a user that has been transmitting with the $u_0$ root and cyclic shift $\kappa _\mathrm{v}$. As a set of $n^{(\mathrm{rep})} n^{(\mathrm{ant})}$ correlations are available, one per antenna and repetition, the output of the correlators are combined to obtain a single decision statistic. That combination can be done in several forms.

\subsubsection{\Acf{PC}}
The power of the correlation or \acf{PDP} of the preamble per antenna and repetition is combined to make the decision as (green block in Fig. \ref{fig:RxorMC}):
\begin{equation}
\label{eq:powercombination}
   \Psi _{u_0 }^{(\mathrm{p})} [k]  =   \sum\limits_{i = 1}^{n^{(\mathrm{ant})}} \sum\limits_{m = 1}^{n^{(\mathrm{rep})}} \left| {\Phi _{u_0 }^{m,i} [k]} \right|^2.
\end{equation}

\subsubsection{\Acf{CC}}
As \ac{PC} loses the phase information in ${\Phi_u^{m,i} [k]}$, a coherent combining of the repetitions is done before evaluating the absolute value, in a mixed form (blue block in Fig. \ref{fig:RxorMC}): 
\begin{equation}
\label{eq:mixedcombination} 
\Psi _{u_0 }^{(\mathrm{c})} [k]   =    \sum\limits_{i = 1}^{n^{(\mathrm{ant})}} \left| \sum\limits_{m = 1}^{n^{(\mathrm{rep})}} {\Phi_{u_{0}}^{m,i} [k]} \right|^2.
\end{equation}
This option takes advantage of correlation in the time domain for channel realizations and of independence in the antenna domain. 

\subsection{Hypothesis test and design criteria}
\label{subsec:Hypo}

The selected combined statistic $\Psi _{u_0 }[k] $ is employed to distinguish between two actual hypotheses:  ${\mathcal{{H}}_1}$ if a user is using  root $u_0$ and cyclic shift $\kappa _\mathrm{v}$; and ${\mathcal{{H}}_0}$ in case of no such \ac{PRACH} transmission. 
Consequently, we have two possible decisions, named $ {\mathcal{\widehat{H}}_0 } $ and ${\mathcal{\widehat{H}}_1 } $, taken as $ \Phi_{u_0} \in Z_0$ or $Z_1$, where $Z_0, Z_1$ are the decision regions to be designed. The performance requirement of \ac{PRACH} for preamble detection is determined by these two parameters:
\begin{itemize}
    \item False alarm ($\mathrm{fa}$): is defined as a conditional total probability of erroneous detection of the preamble when the input is only noise, evaluated as:
    \begin{equation}
    \label{eq:falsealarm}
    p^{(\mathrm{fa})} (Z_1)= \text{Pr} \left\{ {\widehat{\mathcal{H}}_1 |\mathcal{H}_0 } \right\}.
    \end{equation}
    \item True detection ($\mathrm{td}$): is defined as the conditional probability of detection of the preamble when the signal is present:
    \begin{equation}
    \label{eq:detection}
    p^{(\mathrm{td})}(Z_1)  = \mathrm{Pr} \left\{ {\widehat{\mathcal{H}}_1 |\mathcal{H}_1 } \right\}.
    \end{equation}
\end{itemize}
The standard fixes a maximum false alarm probability, $p^{(\mathrm{fa\_des})}$, for specific \ac{PRACH} formats, channel models, and \ac{SNR} points \cite{3gpp38141-1}. Consequently, the design must detect as many device accesses as possible, while keeping the probability of false alarms limited, designing $Z_1$ such as:  
\begin{equation}
\label{eq:criteria}
\begin{gathered}
  \max_{Z_{1}} p^{(\mathrm{td})} (Z_{1})  \hfill \\
  \text{s.t. } p^{(\mathrm{fa})} (Z_{1}) \leqslant p^{(\mathrm{fa\_des})}  \hfill \\ 
\end{gathered}     
\end{equation}
Since we have combined the \ac{PRACH} signal from multiple antennas and sequences to reduce it to a single statistical decision, the decision region $Z_1$ translates into calculating a threshold $\mathcal{T}$, as shown in Fig. \ref{fig:RxorMC}:
\begin{equation}
\label{eq:detectionDecision}
\Psi _{u_0 }^{} [k]\mathop {\mathop  \gtreqqless \limits^{\widehat{\mathcal{H}}_1 } }\limits_{\widehat{\mathcal{H}}_0 } \mathcal{T}.
\end{equation}
Note that there is a trade-off, as attempting to reduce the probability of false alarm will always lead to a deterioration in the probability of detection, and vice versa.

\section{Probabilistic analysis}
\label{sec:transitionProb}
In order to design a detector, the transition probabilities, that is, the distributions for the vector $\Psi_{u_{0}}$ have to be evaluated. It is clear from eq. \eqref{eq:phi_t,i} that $\Phi _{u_0 }^{m,i} [k]$ conditioned on ${\mathbf{h}}$ (explicitly indicated by including ${(\mathbf{h})}$ in the description) are $n^{(\mathrm{ant})}n^{(\mathrm{rep})}$ complex Gaussian random variables, each with its own average and the same variance:
\begin{equation}
\label{eq:distphi}
\Phi _{u_0 }^{m,i} [k]({\mathbf{h}}) \sim {\cal CN}(\mu_{u_0 }^{m,i} [k]({\mathbf{h}}),\sigma _z^2 ). 
\end{equation}
The output of the combiner for a given channel sample would always results in the sum of Gaussian variables with a mean different from 0, which are subsequently squared. It is well known that this squaring yields a scaled non-central $\chi^2$ distribution, commonly notated as $\chi^{'2}$ \cite{patnaik1949non}. The non-centrality parameter $\lambda_{u_0}[k]({\mathbf{h}})$ is distributed as a mix of linear and squared combinations of channel samples. 

As $h_g^{m,i}$ are complex Gaussian variables, their weighted combinations in eq. \eqref{eq:mu_t,i} can be also described as Gaussian after deconditioning on ${\mathbf{h}}$:
\begin{equation}
\label{eq:mudist}
    \mu_{u_0 }^{m,i} [k]({\bf{h}})\sim \mathcal{CN}(0,\sigma_{u_{0}}^{2} [k]),
\end{equation}
with 
\begin{equation}
\label{eq:sigma_epsilon2}
\sigma_{u_{0}}^{2} [k] = \sigma _h^2 \sum\limits_{g \in {\cal D}}^{} {\left| {c_{u_{(g)} ,\mathrm{v}_g ,u_0 ,0}^{\epsilon_g } [k]} \right|^2}.    
\end{equation}

The variance $\sigma_{u_{0}}^{2} [k]$ depends on the set of transmitting devices for each \ac{PRACH} occasion, the roots and the \acp{CS} they are using, as well as the undergoing \acp{CFO}:  
\begin{itemize}
    \item Under $\mathcal{H}_0$, the received signal does not include any device transmission for the root $u_0$, but it might contain interference from other roots; thus: 
    \begin{equation}
    \label{eq:sigma_epsilonH0}
    \hat{\sigma}_{u_{0}}^{2} [k] = \sigma_h^2 \sum\limits_{g \in \bar{\mathcal{D}}_{u_{0}}}{\left| {c_{u_{(g)} ,\mathrm{v}_g ,u_0 ,0}^{\epsilon_g } [k]} \right|^2 },
    \end{equation} 
    \item  Under $\mathcal{H}_1$, the received signal also includes the transmitted signal for \ac{CS} $\kappa _\mathrm{v}$ and root $u_0$; thus, for that specific cyclic shift: 
    \begin{equation}
    \label{eq:sigma_epsilonH1}
    \check{\sigma}_{u_{0}}^{2} [k] = \sigma _h^2 |c_{u_0 ,\mathrm{v} ,u_0 ,0}^{\epsilon_0 } [k]|^2  + \hat{\sigma}_{u_{0}}^{2} [k]. 
    \end{equation}
\end{itemize}

As we obtain the statistics, we will use ${\sigma }_{u_{0}}^2$ as any one of them. 

\subsection{\Acf{PC}}
\label{subsec:pdpPower}

Conditioned on channel samples, all realizations of $\Phi_{u_0}^{m,i}[k]$ are independent; thus, 
$\Psi^{(\mathrm{p})}_{u_0}[k]$ in eq. \eqref{eq:powercombination} is the addition of the absolute value squared of the $n^{(\mathrm{ant})} n^{(\mathrm{rep})}$ variables following the Gaussian distribution in eq. \eqref{eq:distphi}, which is the scaled $\chi^{'2}$ distribution that can be expressed as:
\begin{equation}
\label{eq:psincchi2}
\Psi^{(\mathrm{p})}_{u_0} [k]({\bf{h}}) \sim \sigma _z^2 \chi _{2n^{(\mathrm{ant})} n^{(\mathrm{rep})} }^{'2} \left( {\frac{{\lambda^{(\mathrm{p})}_{u_{0}} [k]({\bf{h}})}}{{\sigma _z^2 }}} \right),
\end{equation}
with the non-centrality parameter given by
\begin{equation}
\label{eq:lambda_pdp}
\lambda^{(\mathrm{p})}_{u_{0}} [k]({\bf{h}}) = \sum\limits_{i = 1}^{n^{(\mathrm{ant})}} {\sum\limits_{m = 1}^{n^{(\mathrm{rep})}} {\left| {\mu _{u_0 }^{m,i} [k]({\bf{h}})} \right|^2 }}.  
\end{equation}

As previously described, two key cases for the distribution of the channel samples have been evaluated:
\subsubsection{Independent channel realizations per repetition}
\label{subsubsec:pdp_indep}
We assume each repetition sees an independent channel realization. Thus, $\lambda^{(\mathrm{p})}_{u_{0}} [k]$ is the addition of the squared absolute value of independent complex Gaussian variables, i.e., a $\chi^2$ distribution: $\lambda^{(\mathrm{p})}_{u_{0}} [k]({\mathbf{h}}) \sim \frac{\sigma_{u_{0}}^2 [k]}{2} \chi_{2n^{(\mathrm{ant})} n^{(\mathrm{rep})} }^2$. 

As detailed in the Appendix, the marginal distribution of a scaled $\chi^{'2}$ distribution (eq. \ref{eq:psincchi2}) with the non-centrality parameter following a scaled $\chi^2$ distribution with the same \ac{DOF} is also a $\chi^2$ distribution with the same \ac{DOF}. Thus, the distribution for combination is:
\begin{equation}
\label{eq:pdp_indep}
 \dot \Psi_{u_0}^{(\mathrm{p})} [k] \sim \sigma_z^2 \left(1  + \frac{\sigma_{u_{0}}^2 [k]}{2\sigma_z^2} \right) \chi^2_{2n^{(\mathrm{ant})}n^{(\mathrm{rep})}}.   
\end{equation} 
with the dot above the variable indicating the distribution under independent channel realizations. 
Its \ac{CCDF} can be expressed following eq. \eqref{eq:CDFofChisquare} of Appendix as:
\begin{equation}
    \label{eq:Pfadot}
    \bar F_{\dot{\Psi }_{u_0}^{(\mathrm{p})} [k] } (\psi) = \frac{\Gamma \left( {n^{(\mathrm{ant})} n^{(\mathrm{rep})} ,\frac{\psi} {{2\sigma _z^2  + {\sigma }_{u_{0}}^2 [k] }}} \right)} {{\Gamma (n^{(\mathrm{ant})} n^{(\mathrm{rep})} )}}.
\end{equation}
It is important to note that $\sigma_{u_{0}}^2 [k]$ can represent $\hat{\sigma}_{u_{0}}^{2} [k]$ or $\check{\sigma}_{u_{0}}^{2} [k]$ depending on whether the \ac{CCDF} is conditioned on $\mathcal{H}_0$ or $\mathcal{H}_1$, respectively.

\subsubsection{Identical channel realizations per repetition}
\label{subsubsec:pdp_equal}
In this other key case, the different repetitions are carried out over identical channel realizations, then the non-centrality parameter can be expressed as:
\begin{equation}
\label{eq:identicalChannelLamba}
\ddot\lambda^{(\mathrm{p})}_{u_{0}} [k]({\bf{h}})  = n^{(\mathrm{rep})} \sum\limits_{i = 1}^{n^{(\mathrm{ant})}} {\left| {\mu _{u_0 }^{1,i} [k]({\bf{h}})} \right|^2 }. 
\end{equation}
As channel realizations between antennas are independent, $\sum\limits_{i = 1}^{n^{(\mathrm{ant})}} {\left| {\mu _{u_0}^{1,i} [k]({\mathbf{h}})} \right|^2} \sim \frac{\sigma_{u_{0}}^2 [k]}{2} \chi_{2n^{(\mathrm{ant})}}^2 $, with $\sigma_{u_{0}}^{2} [k]$ still given by eq. \eqref{eq:sigma_epsilon2}. Therefore, the non-centrality parameter distribution can be described as $\ddot \lambda^{(\mathrm{p})}_{u_{0}} [k]({\mathbf{h}})\sim n^{(\mathrm{rep})} \frac{\sigma_{u_{0}}^2 [k]}{2} \chi _{2n^{(\mathrm{ant})}}^2$. 

Again, we have the $\chi^{'2}$ in eq. \eqref{eq:psincchi2} conditioned on a $\chi^2$ distribution but now the \ac{DOF} are different for both variables. Thus, as shown in the Appendix, the distribution can be written as the convolution of two $\chi^2$ distributions: 
\begin{equation}
\label{eq:pdp_indep2}
 \ddot \Psi_{u_0}^{(\text{p})} [k] \sim \sigma _z^2 \chi _{2n^{(\mathrm{ant})} (n^{(\text{rep})}  - 1)}^2 * \sigma_z^2 \left(1  + \frac{n^{(\text{rep})} \sigma_{u_{0}}^2 [k]}{2\sigma_z^2} \right)\chi _{2n^{(\text{ant})} }^2.
\end{equation} 
where $*$ is the convolution operator and the double dot above the variable indicates the distribution under identical channel realizations.

Consequently, its \ac{CCDF} can be expressed following eq. \eqref{eq:CDFofX} of Appendix:
\begin{multline} \label{eq:CCDF_iden_power_comb}
    \bar F_{\ddot{\Psi }_{u_0}^{(\mathrm{p})} [k] } (\psi) = 1 - \sum_{j=0}^{\infty} \frac{\Gamma(j+n^{(\mathrm{ant})})}{j!\Gamma(n^{(\mathrm{ant})})} \left( \frac{\sigma^{2}_{\theta} [k]}{\sigma^{2}_{\theta} [k] +1} \right)^{j} \\
     \left( \frac{1}{\sigma^{2}_{\theta} [k] + 1} \right)^{n^{(\mathrm{ant})}}  \frac{\gamma \left( n^{(\mathrm{ant})}n^{(\mathrm{rep})}+j, \frac{\psi}{2\sigma _z^2} \right)}{\Gamma(n^{(\mathrm{ant})}n^{(\mathrm{rep})}+j)},
\end{multline}
where $\sigma^{2}_{\theta} [k] = n^{(\mathrm{rep})}\sigma_{u_{0}}^2 [k] / 2\sigma^{2}_{z}$.

\subsection{\Acf{CC}}
\label{subsec:coh2}

This scheme performs a coherent combining of the repetitions before evaluating the absolute value, to subsequently combine by antenna. In this case, the inner sum of the absolute value of eq. (\ref{eq:mixedcombination}) for each antenna follows the next distribution:
\begin{equation}
\label{eq:innerSummationCC}
    \sum\limits_{m = 1}^{n^{(\text{rep})}} \Phi^{m,i}_{u_{0}} [k] ({\mathbf{h}}) \sim \mathcal{CN}\left({ {\sum\limits_{m = 1}^{n^{(\text{rep})}} {\mu _{u_0 }^{m,i} [k]({\mathbf{h}})} } , n^{(\text{rep})} \sigma _z^2 }  \right).
\end{equation}
As before, conditioned on channel samples, all the antenna channel realizations are independent; then, $\Psi^{(\mathrm{c})}_{u_{0}}[k]$ in eq. \eqref{eq:mixedcombination} is the addition of the square of the absolute value of $n^{(\mathrm{ant})}$ variables following the Gaussian distribution in eq. \eqref{eq:innerSummationCC}, which is the following scaled $\chi^{'2}$ distribution:
\begin{equation}
    \Psi^{(\mathrm{c})}_{u_{0}}[k]\sim  n^{(\mathrm{rep})} \sigma _z^2 \chi_{2n^{(\mathrm{ant})}}^{'2} \left(\frac{\lambda^{(\mathrm{c})}_{u_{0}} [k]({\mathbf{h}})}{n^{(\mathrm{rep})} \sigma _z^2} \right),
\end{equation}
with non-centrality parameter:
\begin{equation}
\label{eq:non-centralityCC}
    \lambda^{(\mathrm{c})}_{u_{0}} [k]({\mathbf{h}}) = \sum\limits_{i = 1}^{n^{(\mathrm{ant})}} \left| { {\sum\limits_{m = 1}^{n^{(\mathrm{rep})}} {\mu _{u_0 }^{m,i} [k]({\mathbf{h}})} } } \right|^2.
\end{equation}

Next, the same two key cases have been evaluated.
\subsubsection{Independent channel realizations per repetition}
\label{subsubsec:coh_indep}
Now, the additions within $\dot \lambda^{(\mathrm{c})}_{u_{0}} [k]$ in eq. \eqref{eq:non-centralityCC} are all independent; thus, the marginal distribution of the inner sum for each antenna is given by: $ {\sum\limits_{m = 1}^{n^{(\mathrm{rep})}} {\mu _{u_0 }^{m,i} [k]({\mathbf{h}})} } \sim \mathcal{CN}(0, n^{(\mathrm{rep})} \sigma_{u_{0}}^2 [k])$.
Therefore, $\dot \lambda^{(\mathrm{c})}_{u_{0}} [k]$ is distributed as the addition of $2n^{(\mathrm{ant})}$ Gaussian variables with 0 mean and $ n^{(\mathrm{rep})} \sigma_{u_{0}}^2 [k]$ variance; thus, $\dot \lambda^{(\mathrm{c})}_{u_{0}} [k]\sim  n^{(\mathrm{rep})} \frac{\sigma_{u_{0}}^2 [k]}{2} \chi_{2n^{(\mathrm{ant})}}^2$.  

Consequently, we have again a scaled $\chi^{'2}$ distribution with a scaled  $\chi^2$ as a non-centrality parameter, both with two \ac{DOF}. Therefore, the marginal distribution can be expressed as: 
\begin{equation}
\label{eq:PsiIndepChannelHyb}
\dot\Psi _{u_0 }^{(c)} [k] \sim  n^{(\mathrm{rep})}\sigma _{z}^2 \left(1  + \frac{\sigma_{u_{0}}^2 [k]}{2\sigma _{z}^2} \right)\chi_{2n^{(\mathrm{ant})}}^2,
\end{equation}
with its corresponding \ac{CCDF} as eq. \eqref{eq:CDFofChisquare} of Appendix: 
\begin{equation}
\label{eq:PsiIndepChannelHybCCDF}
\bar {F}_{\dot \Psi_{u_0}^{(\mathrm{c})}[k]} (\psi) = \frac{\Gamma \left( {n^{(\mathrm{ant})} ,\frac{\psi} {{2 n^{(\mathrm{rep})} \sigma _z^2  + n^{(\mathrm{rep})} {\sigma }_{u_{0}}^2 [k]}}} \right)}{\Gamma \left( n^{(\mathrm{ant})} \right)}.
\end{equation}

\subsubsection{Identical channel realizations per repetition}
\label{subsubsec:coh_equal}
Now, as the different repetitions has the same channel realization, the non-centrality parameter in eq. \eqref{eq:non-centralityCC} can be expressed as ${\ddot \lambda}^{(\mathrm{c})}_{u_{0}} [k]({\mathbf{h}}) = \sum\limits_{i = 1}^{n^{(\mathrm{ant})}} \left| {n^{(\mathrm{rep})}  {\mu _{u_0 }^i [k]({\mathbf{h}})} } \right|^2$,
which, since channel realizations between antennas are independent, follows the distribution $\ddot{\lambda}^{(\mathrm{c})}_{u_{0}} [k]({\mathbf{h}}) \sim \left( {n^{(\mathrm{rep})} } \right)^2  \frac{\sigma_{u_{0}}^2 [k]}{2} \chi_{2n^{(\mathrm{ant})}}^2$. 

Then, $\ddot \Psi_{u_0 }^{(\mathrm{c})} [k]$ is a scaled $\chi^{'2}$ with a scaled $\chi^2$ both with two \ac{DOF}, that as we have seen before, can be expressed as: 
\begin{equation}
\label{eq:PsiEqualChannelHyb}
\ddot \Psi_{u_0 }^{(\mathrm{c})} [k] \sim  n^{(\mathrm{rep})} \sigma _z^2 \left(1  + \frac{n^{(\mathrm{rep})} \sigma_{u_{0}}^2 [k]}{2\sigma _z^2}  \right)\chi_{2n^{(\mathrm{ant})}}^2,
\end{equation}
with \ac{CCDF}: 
\begin{equation}
\label{eq:PsiEqualChannelHybCCDF}
\bar {F}_{\ddot \Psi_{u_0}^{(\mathrm{c})}[k]} (\psi) = \frac{\Gamma \left( {n^{(\mathrm{ant})} ,\frac{\psi} {{2 n^{(\mathrm{rep})} \sigma _z^2  +\left( {n^{(\mathrm{rep})} } \right)^2 {\sigma }_{u_{0}}^2 [k]}}} \right)}{\Gamma \left( n^{(\mathrm{ant})}\right)}.
\end{equation}

\section{Optimal detector without CFO}
\label{sec:OptimalDetectorWithoutCFO}
In this section, we define the optimal detector for the absence of \ac{CFO}. This is based on the premise that the correlation of a sequence with the same root as the reference is zero, except at the \ac{CS} of the transmitted device $\kappa_{\mathrm{v}_g}$, since there is no power transfer between correlation lags $k$, while the effects of other roots can be considered as an increase in the noise level. 

Considering this case, the correlation values simplify to those of eq. \eqref{eq:corrSameRoot} and \eqref{eq:corrDiffRoot}, while $\sigma_{u_{0}}^{2}$ can be expressed as:  
\begin{itemize}
    \item Under $\mathcal{H}_0$:
    \begin{equation}
    \label{eq:sigmaH0noCFO}
    \hat{\sigma}_{u_{0}}^{2}  = \sigma _h^2 \sum_{g \in \bar{\mathcal{D}}_{u_{0}}} \frac{1}{{\ell}} = \sigma _h^2 \frac{n_{u_{0}}^{(\mathrm{inter})}}{{\ell}},
    \end{equation}
    \item  Under $\mathcal{H}_1$: 
    \begin{equation}
    \label{eq:sigmaH1noCFO}
    \check{\sigma}_{u_{0}}^{2}  = \sigma _h^2 \left( {1 + \frac{{n_{u_{0}}^{(\mathrm{inter})} }} {{\ell }}} \right).
    \end{equation}
\end{itemize}
where $n_{u_{0}}^{(\mathrm{inter})}  = n^{(\mathrm{dev})}-n^{(\mathrm{dev})}_{u_0}$ is the number of devices transmitting using a different root than the reference. Note that $\hat{\sigma}_{u_{0}}^{2}$ and $\check{\sigma}_{u_{0}}^{2}$ do not depend on the lag $k$, simplifying the analysis as all lags exhibit the same variance.


A false alarm might occur with probability $p_{u_0}^{(\mathrm{fa})} [k] (\mathcal{T}) =  \Pr \left\{ {\Psi _{u_0 }[k] > \mathcal{T}|\mathcal{H}_0 } \right\} = \bar F_{\Psi_{u_0} [k]|\mathcal{H}_0 } (\mathcal{T})$, that is, the false alarm is given by the \ac{CCDF} for  $\Psi _{u_0} [k]$ conditioned on no \ac{PRACH} transmission for the root and \ac{CS} under test. 
On the other hand, a true detection would happen with probability $p_{u_0}^{(\mathrm{td})} [k](\mathcal{T})= \Pr \left\{ {\Psi _{u_0 } [k] > \mathcal{T}|\mathcal{H}_1 } \right\} =\bar F_{\Psi _{u_0 } [k]|\mathcal{H}_1 } (\mathcal{T}).$ 

As a design criterion, assuming that the devices on each \ac{PRACH} occasion randomly select \acp{CS} among all configured roots, we can assume that the mean of $n^{(\mathrm{inter})}_{u_{0}}$ is the same for all roots. Furthermore, since the channel is flat and the circular correlation between sequences with the same root is a Kronecker delta, while that between different roots is a constant, we can consider all roots and cyclic shift as equivalent. In this way, the same threshold can be used for all roots configured in the \ac{gNB}. A false alarm might occur in any root, thus, in order to keep the false alarm rate below the maximum, $p^{(\mathrm{fa})}(\mathcal{T}) \leq p^{(\mathrm{fa\_des})}$, the test for each root and correlation lag $k$ should be kept below a value $p^{(\mathrm{fa\_sample\_des})}$, which is the desired probability of false alarm per lag and per root, easily obtained assuming independence among samples. Consequently, we can obtain the optimal threshold for both signal combiners solving the following equation:
\begin{equation}
\label{eq:ThrEqDetermination}
p^{(\mathrm{fa\_sample\_des})} = 1 -  \left( {1 - p^{(\mathrm{fa\_des})}}\right)^{\frac{1}{n^{(\text{root})}\ell}} = \bar F_{\Psi_{u_0} [k]|\mathcal{H}_0 } (\mathcal{T}).
\end{equation}



\subsection{Optimal threshold for \ac{PC}}
\label{subsec:pdp}

\subsubsection{Independent channel realizations per repetition}
\label{subsubsec:pdp_indep2}

As described above, the optimal threshold is obtained by solving eq. \eqref{eq:ThrEqDetermination}, using the \ac{CCDF} described in eq. \eqref{eq:Pfadot} conditioned on the $\mathcal{H}_0$ hypothesis, that is, with $\hat{\sigma}_{u_{0}}^{2}$ defined in eq. \eqref{eq:sigmaH0noCFO}:
\begin{equation} \label{eq:Thres_p_comb}
\dot{\mathcal{T}}_{u_{0}}^{(\mathrm{p})}  = \left( 2\sigma _z^2  + \hat{\sigma}_{u_{0}}^{2} \right) \Gamma ^{(\mathrm{inv})} \left( {n^{({\mathrm{ant}})} n^{(\mathrm{rep})} ,p^{(\mathrm{fa\_sample\_des})}} \right),
\end{equation}
where $\Gamma^{(\mathrm{inv})}$ is the inverse of the upper incomplete Gamma function. 

Given that $\hat{\sigma}_{u_{0}}^{2}$ depends on $n^{(\mathrm{inter})}_{u_{0}}$, $\dot{\mathcal{T}}_{u_{0}}^{(\mathrm{p})}$ is an instantaneous threshold whose value would have to be updated in each \ac{PRACH} occasion. Since $n^{(\mathrm{inter})}_{u_{0}}$ is unknown a priori, the mean of its distribution based on cell traffic could be used instead, in which case we would have an average threshold to achieve the target false alarm over a period of time. This approach is also valid, as the standard dictates that it must be statistically fulfilled.

Following, the true detection probability can also be obtained under $\mathcal{H}_1$ hypothesis plugging the calculated threshold in eq. \eqref{eq:Pfadot} while the variance is $\check{\sigma}_{u_{0}}^{2}$ defined in eq. \eqref{eq:sigmaH1noCFO}: 
\begin{equation}
\label{eq:probDetectionPowerCombIndepRep}
    \dot p^{(\mathrm{td}) (\mathrm{p})} [\dot{\mathcal{T}}_{u_{0}}^{(\mathrm{p})}]= \frac{\Gamma \left( {n^{({\mathrm{ant}})} n^{(\mathrm{rep})} ,\frac{\dot{\mathcal{T}}_{u_{0}}^{(\mathrm{p})}}
{{2{\sigma } _z^2  + \check{\sigma}_{u_{0}}^{2} }}} \right)} {{\Gamma \left(n^{({\mathrm{ant}})} n^{(\mathrm{rep})} \right)}}.
\end{equation}

\subsubsection{Identical channel realizations per repetition}
\label{subsubsec:pdp_equal2}
The threshold, $\ddot{\mathcal{T}}_{u_{0}}^{(\mathrm{p})}$, can be obtained from the \ac{CCDF} in eq. (\ref{eq:CCDF_iden_power_comb}) under the $\mathcal{H}_0$ hypothesis. In this case, there is no closed-form for the optimal threshold, since the function has no inverse, but the value is easily obtained with numerical methods as the \ac{CCDF} is monotonically growing. Finally, the detection probability can be expressed as: $\ddot{p}^{(\mathrm{td}) (\mathrm{p})} [\ddot{\mathcal{T}}_{u_{0}}^{(\mathrm{p})}] = \bar F_{\ddot{{\Psi }}_{u_0}^{(\mathrm{p})} [k]|\mathcal{H}_1 } (\ddot{\mathcal{T}}_{u_{0}}^{(\mathrm{p})})$.

\subsection{Optimal threshold for \ac{CC}}
\label{subsec:coh}

\subsubsection{Independent channel realizations per repetition}
\label{subsubsec:coh_indep2}
Following the same steps as before, the optimal threshold is determined solving eq. \eqref{eq:ThrEqDetermination}, using the CCDF in eq. \eqref{eq:PsiEqualChannelHybCCDF} conditioned on $\mathcal{H}_0$ hypothesis:
\begin{equation*}
\dot{\mathcal{T}}_{u_{0}}^{(\mathrm{c})}  =  n^{(\mathrm{rep})} \left(2 \sigma _z^2  +  \hat{\sigma}_{u_{0}}^{2}\right) \Gamma ^{(\text{inv})} \left( {n^{({\mathrm{ant}})} ,p^{(\mathrm{fa\_sample\_des})}} \right).
\end{equation*}
On the other hand, the detection probability can be obtained as $\dot{p}^{(\text{td}) (\mathrm{c})}[\dot{\mathcal{T}}_{u_{0}}^{(\mathrm{c})}]= \bar {F}_{\dot \Psi_{u_0}^{(\mathrm{c})}[k] |\mathcal{H}_1} (\dot{\mathcal{T}}_{u_{0}}^{(\mathrm{c})})$ under $\mathcal{H}_1$ hypothesis.

\subsubsection{Identical channel realizations per repetition}
\label{subsubsec:coh_equal2}

In a similar way to the previous case, the optimal threshold would be:
\begin{multline*}
\ddot{\mathcal{T}}_{u_{0}}^{(\text{c})}  = \\ n^{(\mathrm{rep})} \left(2 \sigma _z^2  +  n^{(\mathrm{rep})} \hat{\sigma}_{u_{0}}^{2}\right) \Gamma ^{(\text{inv})} \left( {n^{({\text{ant}})} ,p^{(\mathrm{fa\_sample\_des})}} \right),
\end{multline*}
and the corresponding detection probability computed as $\ddot p^{(\mathrm{td}) (\mathrm{c})} [\ddot{\mathcal{T}}_{u_{0}}^{(\mathrm{c})}]= \bar {F}_{\ddot \Psi_{u_0}^{(\mathrm{c})} [k]|\mathcal{H}_1} (\ddot{\mathcal{T}}_{u_{0}}^{(\mathrm{c})})$.

\section{Detector design with \ac{CFO}}
\label{sec:DetectorDesignCFO}
We modify the detection algorithm to include the possibility of \ac{CFO}. We assume \ac{PC} and independent channel realizations per repetition for simplicity, as \ac{CC} and equal channel realizations can be added as in the previous analysis.

\begin{figure}[!ht]
\begin{center}
\includegraphics[width=\columnwidth]{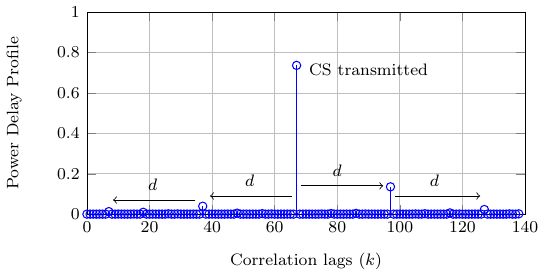}
\caption{Format C0 PDP with $\epsilon=0.3$ and $u_{0}=51$ without noise and channel.}
\label{fig:CFO_PDP}
\end{center}
\end{figure}

When the signal is affected by \ac{CFO}, part of the signal power is spread over all correlation lags, creating non-uniform intra-root interference. In the case of $\epsilon < 0.5$, this interference is distributed in descending order across the lags, maintaining a constant distance between them with respect to the transmitted preamble. Fig. \ref{fig:CFO_PDP} shows an example without noise or channel effects of format C0 with $\epsilon=0.3$ and reference root $u_{0}=51$. This inter-sample distance can be calculated as \cite{wang2015multiuser}:
\begin{equation}
\label{eq:inter-sampleDistance}
d = \left\{ {\begin{array}{*{20}c}
   {d_u } \hfill & {{\text{ if }}0 \leqslant d_u  < \ell /2} \hfill  \\
   {\ell - d_u } \hfill & {{\text{ if }}d_u  \geqslant \ell /2} \hfill  \\
 \end{array} } \right.
\end{equation}
where $d_u$ is the smallest non-negative integer satisfying $(u \cdot d_u)_\ell = 1$. It can be seen that the true peak produces false peaks with a smaller amplitude as the distance increases. Consequently, the lag order from largest to smallest \ac{PDP} amplitude is $(\kappa_{\text{v}} + b \cdot d)_\ell$, where $b\in \mathbb{Z}$. 


In the presence of channel and noise, it may happen that one of these false peaks has a higher amplitude than the true peak. Therefore, the possible events that can trigger a false alarm if the above detector is used are:
\begin{itemize}
    \item Noise plus interference from other roots causes a sample exceed the threshold $\mathcal{T}$.
    \item If there is a true peak at $k^{(\mathrm{t})}$ which gives rise to a false peak at $k^{(\mathrm{}{f})}=(k^{(\mathrm{t})} + b \cdot d)_{\ell}$, which is greater than $\mathcal{T}$. This produces a false alarm in $k^{(\mathrm{f})}$ whether or not $k^{(\mathrm{t})}$ is detected. 
\end{itemize}

This behavior makes the distribution of $\Psi$ dependent on the lag $k$, and hence on the threshold. As a consequence, the \ac{CCDF} of $\Psi_{u_{0}}[k]$ for \ac{PC} follows eq. (\ref{eq:Pfadot}) and its threshold can be expressed for the reference root $u_{0}$ as:
\begin{equation}
\label{eq:ThresholdCFODetector}
\dot{\mathcal{T}}^{(\mathrm{CFO})}_{u_{0}} [k] = \frac{\Gamma ^{(\mathrm{inv})} \left( {n^{({\mathrm{ant}})} n^{(\mathrm{rep})} ,p^{(\mathrm{fa\_sample\_des})}} \right)}{\left( 2\sigma _z^2  + {\sigma}_{u_{0}}^{2} [k] \right)^{-1}},
\end{equation}
where ${\sigma}_{u_{0}}^{2} [k]$ follows eq. (\ref{eq:sigma_epsilon2}).

As above, the total probability of false alarm will be:
\begin{equation}
\label{eq:TotalpFACFOdetector}
    \dot p^{({\mathrm{fa}})}  = 1- \prod_{u \in \mathcal{R}} \prod_{k=0}^{\ell-1} \left(1-\bar F_{\dot \Psi _{u } [k]} (\dot{\mathcal{T}}^{(\mathrm{CFO})}_{u} [k]) \right),
\end{equation}
and the true detection probability of the device transmitting at lag $k^{(\mathrm{t})}$:
\begin{equation}
\label{eq:probDetectionCFO}
     \dot p^{(\mathrm{td}) (\mathrm{p})}[k^{(\mathrm{t})}] = \bar F_{\dot \Psi _{u_0 }^{} [k^{(\mathrm{t})}]} (\dot{\mathcal{T}}^{(\mathrm{CFO})}_{u_{0}} [k^{(\mathrm{t})}]).
\end{equation}

\subsection{Detection algorithm with \ac{CFO}}
\label{subsec:intersample}
Since it is not possible to know in advance which devices will transmit on which \ac{CS}, the algorithm must be able to locate them and try to avoid selecting false alarms as true peaks. For this purpose, the following algorithm is proposed: If a set of \acp{CS}, $\mathbf{k}_{s}$, are cyclically separated from each other by $b \cdot d$ samples and exceed the base threshold, $\dot{\mathcal{T}}_{u_{0}}^{(\mathrm{p})}$, only the highest of them is selected as a true peak and the rest are discarded as false ones. The underlying reasoning takes into account that the highest peak is more likely to be the true one. This process is performed for all the configured roots. In addition, the selected peaks also have to exceed the threshold due to \ac{CFO}, $\dot{\mathcal{T}}^{(\mathrm{CFO})}_{u_{0}} [k]$, to ensure that they are not originated due to interference from other roots. The complete algorithm is explained in Algorithm \ref{algo_1}.



\begin{algorithm}[ht] 
  \caption{Peak Detection and Classification} \label{algo_1}
  \begin{algorithmic}[1]
    \ENSURE Classified true peaks for all roots.
    \FOR{each root $u_0 \in \mathcal{R}$}
      \STATE Compute the base threshold assuming no \ac{CFO} as eq. (\ref{eq:Thres_p_comb}) \\
      $\displaystyle \dot{\mathcal{T}}_{u_{0}}^{(\text{p})}  \gets \left( 2\sigma _z^2  + \hat{\sigma}_{u_{0}}^{2} \right) \Gamma ^{(\text{inv})} \left( {n^{({\text{ant}})} n^{(\text{rep})} ,p^{(\text{fa\_sample\_des})}} \right)$
      \STATE Find all candidate peak indices \\
      $\mathbf{k}_{u_0} \gets \{\,k \mid \Psi_{u_0}[k] > \dot{\mathcal{T}}_{u_{0}}^{(\text{p})}\}$
      \STATE Initialize classified set $\mathcal{C}\gets\emptyset$
      \WHILE{$\mathbf{k}_{u_0}\setminus\mathcal{C}\neq\emptyset$}
        \STATE Select largest remaining peak \\
        $\displaystyle k^{(\mathrm{t})}_s \gets \mathop {\arg \max }\limits_{k\in \mathbf{k}_{u_0}\setminus\mathcal{C}} \{\Psi_{u_0}[k] \}$
        \STATE Group neighboring peaks at distance $b\cdot d$ \\
        $\mathbf{k}_s \gets \bigl\{\,(k^{(\mathrm{t})}_s +  b\cdot d)_{\ell}\;\big|\; \Psi_{u_0}[\,(k^{(\mathrm{t})}_s + b\cdot d)_{\ell}] > \dot{\mathcal{T}}_{u_{0}}^{(\text{p})}\bigr\}$
        \STATE Add grouped peaks to classified set:
        $\mathcal{C}\gets \mathcal{C}\,\cup\,\mathbf{k}_s$
      \ENDWHILE
    \ENDFOR
      \STATE \COMMENT{Interference check}
      \FOR{each $k^{(\mathrm{t})}_s$}
        \STATE Compute interference-aware threshold with \ac{CFO}, $\dot{\mathcal{T}}^{(\mathrm{CFO})}_{u_{(s)}}[k^{(\mathrm{t})}_s]$, based on other true detections
        \IF{$\Psi_{u_{(s)}}[k^{(\mathrm{t})}_s] < \dot{\mathcal{T}}^{(\mathrm{CFO})}_{u_{(s)}}[k^{(\mathrm{t})}_s]$}
          \STATE Discard $k^{(\mathrm{t})}_s$
        \ENDIF
      \ENDFOR
  \end{algorithmic}
\end{algorithm}


Using this algorithm, the probability of false alarm will come from the probability of the noise exceeding the threshold and selecting as true detection a false peak originated from the \ac{CFO}. The probability of selecting a false peak instead of a true one is basically affected by the first false peak, i.e., $k^{(\mathrm{f})}=(k^{(\mathrm{t})} + d)_{\ell}$, as it is the highest, and can be expressed for a device $g$ transmitting at the reference root $u_{0}$ as:
\begin{multline*}
\label{eq:pfaDueCFO}
    p^{(\mathrm{fa\_CFO})}_g  = \\
    \Pr \left\{ {\Psi_{u_0 } [k^{(\text{f})}] ({\bf{h}})> \Psi_{u_0 } [k^{(\text{t})} ] ({\bf{h}}) | \Psi_{u_0 } [k^{(\text{f})}] ({\bf{h}}) \geq \mathcal{\tilde{T}}_{u_{0}}} [k^{(\text{f})}] \right\}
\end{multline*}
where $\mathcal{\tilde{T}}_{u_{0}} [k]= \mathrm{max}(\dot{\mathcal{T}}_{u_{0}}^{(\mathrm{p})}, \dot{\mathcal{T}}^{(\mathrm{CFO})}_{u_0}[k])$ since if the false peak is selected, then it also has to exceed the threshold due to \ac{CFO}; and $\Psi_{u_0 } [k^{(\mathrm{f})}] ({\bf{h}})$ along with $\Psi_{u_0 } [k^{(\mathrm{t})}] ({\bf{h}})$ follows the distribution of eq. (\ref{eq:psincchi2}).

This probability can be evaluated as the sum of two event probabilities. The first, $p_g^{(1)}$, is the probability that the false peak is higher than the true peak when both peaks exceed the base threshold, so both are detected and the higher one is selected. The second one, $p_g^{(2)}$, is the probability that the false peak exceeds the threshold while the true peak does not (only the false one is selected):
\begin{equation}
\label{eq:pfaDueCFOcomposite}
    p^{(\mathrm{fa\_CFO})}_g  = p_g^{(1)}+p_g^{(2)},
\end{equation}
where $p_g^{(1)}$ can be computed as:
\begin{equation*}
\label{eq:p_1}
 \int_{\dot{\mathcal{T}}_{u_{0}}^{(\text{p})}}^\infty \underbrace{\int  \int^{\infty}_{-\infty}}_{\#{\bf{h}} }  {\bar F_{\Psi _{u_0 } [k^{(\text{f})}] ({\bf{h}})}  (\psi) f_{\Psi _{u_0 }^{} [k^{(\text{t})} ] ({\bf{h}})} (\psi) f_{\text{H}}({\bf{h}}) d{\bf{h}} d\psi },    
\end{equation*}
and $p_g^{(2)}$ as 
\begin{equation*}
\label{eq:p_2}
 \underbrace{\int  \int^{\infty}_{-\infty}}_{\#{\bf{h}} } \bar F_{\Psi _{u_0 } [k^{(\text{f})}] ({\bf{h}})} (\mathcal{\tilde{T}}_{u_{0}} [k^{(\text{f})}])  F_{\Psi _{u_0 } [k^{(\text{t})} ] ({\bf{h}})} (\dot{\mathcal{T}}_{u_{0}}^{(\text{p})}) f_{\text{H}}({\bf{h}}) d{\bf{h}},
\end{equation*}
being $\#{\bf{h}}=n^{(\mathrm{rep})} n^{(\mathrm{ant})}$ the number of elements of $\bf{h}$.

Consequently, the total false alarm probability can be expressed as:
\begin{equation}
\label{eq:pfaTotalCFOdetector}
    p^{(\mathrm{fa})} = p^{(\mathrm{noise\_plus\_inter})} + \sum_{g \in \mathcal{D}} p^{(\mathrm{fa\_CFO})}_g,
\end{equation}
where $p^{(\mathrm{noise\_plus\_inter})}$ takes into account that the selected peaks have to exceed both the base threshold and the \ac{CFO} threshold. It can be expressed following eq. \eqref{eq:TotalpFACFOdetector} as:
\begin{equation}
\label{eq:probNoisePlusInter}
    p^{(\mathrm{noise\_plus\_inter})} = 1- \prod_{u \in \mathcal{R}} \prod_{k=0}^{\ell-1} \left(1 -  \bar F_{ \dot\Psi _{u } [k]} (\mathcal{\tilde{T}}_{u} [k]) \right).
\end{equation}
Moreover, the true detection probability for the device $g$ can be computed as:
\begin{equation}
\label{eq:probDetectionCFOdetector}
    p^{(\mathrm{td}) (\mathrm{p})}_{g}[k^{(\mathrm{t})}_{g}] = \bar F_{\dot \Psi _{u_0 } [k_{g}^{(\mathrm{t})}]} (\mathcal{\tilde{T}}_{u_{0}} [k_{g}^{(\mathrm{t})}]).
\end{equation}

Knowing the theoretical behavior of the detector allows us to adapt the base threshold, compensating for the CFO effect, and achieve the desired false alarm probability without having a large impact on the detection probability.

\begin{table}
\centering
\begin{threeparttable}
\renewcommand{\arraystretch}{1.3}
\caption{\ac{PRACH} format parameters \cite{3gpp38211}}
\label{tab:Format parameters}
\begin{tabular}{ l l l l l l}
\toprule
\multicolumn{1}{c}{Format} & \multicolumn{1}{c}{$\ell$} & \multicolumn{1}{c}{\ac{PRACH} \ac{SCS}} & \multicolumn{1}{c}{$n^{(\mathrm{rep})}$} & \multicolumn{1}{c}{Bandwidth} & \multicolumn{1}{c}{Duration} \\
\toprule
\multicolumn{1}{c}{0} & \multicolumn{1}{c}{839} & \multicolumn{1}{c}{1.25 kHz} & \multicolumn{1}{c}{1} & \multicolumn{1}{c}{1.08 MHz} & \multicolumn{1}{c}{1 ms}\\
\hline
\multicolumn{1}{c}{C0} & \multicolumn{1}{c}{139} & \multicolumn{1}{c}{$15\cdot 2^\mu$ kHz} & \multicolumn{1}{c}{1} & \multicolumn{1}{c}{$2.16 \cdot 2^{\mu}$ MHz} & \multicolumn{1}{c}{0.144 ms} \\
\hline
\multicolumn{1}{c}{B1} & \multicolumn{1}{c}{139} & \multicolumn{1}{c}{$15\cdot 2^\mu$ kHz} & \multicolumn{1}{c}{2} & \multicolumn{1}{c}{$2.16 \cdot 2^{\mu}$ MHz} & \multicolumn{1}{c}{0.142 ms}\\
\hline
\multicolumn{1}{c}{B2} & \multicolumn{1}{c}{139} & \multicolumn{1}{c}{$15\cdot 2^\mu$ kHz} & \multicolumn{1}{c}{4} & \multicolumn{1}{c}{$2.16 \cdot 2^{\mu}$ MHz} & \multicolumn{1}{c}{0.285 ms}\\
\hline
\multicolumn{1}{c}{B4} & \multicolumn{1}{c}{139} & \multicolumn{1}{c}{$15\cdot 2^\mu$ kHz} & \multicolumn{1}{c}{12} & \multicolumn{1}{c}{$2.16 \cdot 2^{\mu}$ MHz} & \multicolumn{1}{c}{0.856 ms}\\
\bottomrule
\end{tabular}
\begin{tablenotes}
      \item Note: $\mu \in \{0,1,2,3\}$ is the numerology parameter.
    \end{tablenotes}
\end{threeparttable}
\end{table}

\vspace{-0.25cm}
\section{Numerical results}
\label{sec:NumericalSimResults}

In this section, we evaluate the theoretical expressions derived above for different formats supported in 5G \ac{NR} as summarized in Table \ref{tab:Format parameters}. All are configured with a 15~kHz \ac{SCS}, except format 0, which only supports 1.25~kHz, and spread over $\ell$ subcarriers; thus, all \ac{PRACH} formats can be assumed to be located within their coherence bandwidth for delay spreads less than or equal to 100 ns. Furthermore, the coherence time for all formats can be guaranteed for frequency Doppler of up to 500~Hz, since their durations are less than 1 ms. In these scenarios, we can assume that the channel realizations per repetition are approximately equal. Additionally, the number of roots configured in \ac{gNB} are $n^{(\mathrm{root})}=2$ and the desired maximum $p^{(\mathrm{fa\_des})}$ is set to $10^{-3}$ to determine the optimal threshold for each combination technique. 

The closed-form expressions obtained in the previous section have been validated and verified with Monte Carlo simulations. Empirical results were obtained using $6\cdot10^{5}$ \ac{PRACH} occasions. 

\subsection{Signal combiners comparison without \ac{CFO}}
\label{subsec:ResultDetProbDiffComb}

\begin{figure}[t]
\begin{center}
\includegraphics[width=\columnwidth]{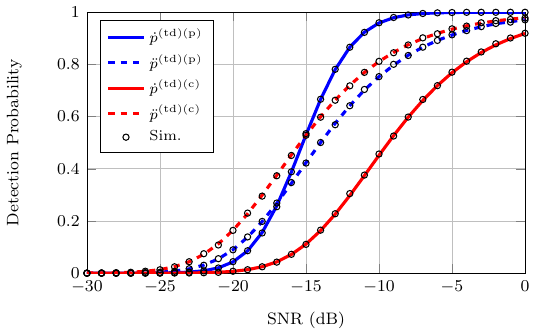}
\caption{Detection probability comparison of combining techniques for format B1 with $n^{(\mathrm{ant})}=1$ and $n^{(\mathrm{inter})}=0$.}
\label{fig:5}
\end{center}
\end{figure}

We first consider the no \ac{CFO} case to compare the performance of different signal combiners. A comparison of the different combination techniques can be seen in Fig. \ref{fig:5} as a function of the \ac{SNR} per antenna, for B1 format with $n^{(\mathrm{ant})}=1$ and $n^{(\mathrm{inter})}=0$. Each combiner computes its optimal threshold based on previous development and measures the detection probability. We observe that \ac{CC} is the best strategy in the case of identical channel realizations, while \ac{PC} outperforms \ac{CC} in the case of independent channel realizations. It is worth noting that, in the low-\ac{SNR} regime, combiners assuming identical channel realizations perform better than independent ones. This aspect should be considered when designing a low-\ac{SNR} detector, which will interfere less with other base stations or other configured roots.


\begin{figure}[!ht]
\begin{center}
\includegraphics[width=\columnwidth]{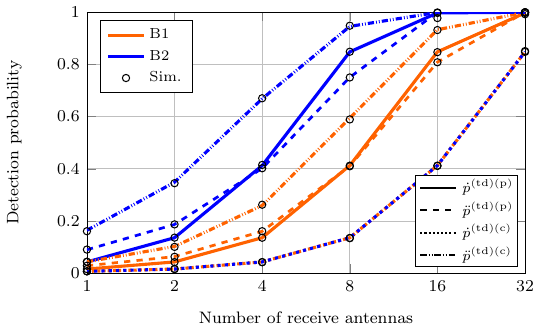}
\caption{Detection probability comparison between formats B1 and B2 for different number of receive antennas, \ac{SNR} of -20 dB, and $n^{(\mathrm{inter})}=0$.}
\label{fig:4}
\end{center}
\end{figure}

Performance also depends on the number of receiving antennas and repetitions. In Fig. \ref{fig:4} we evaluate and compare the performance of the combiners using formats B1 and B2 for a \ac{SNR} of -20 dB and $n^{(\mathrm{inter})}=0$. We observe that format B2, which has twice as many repetitions, outperforms format B1 for all combiners, except in the case of \ac{CC} with independent channel realizations, since in this case the array gain obtained by increasing the number of repetitions is zero. Furthermore, as mentioned above, when dealing with a low-\ac{SNR}, \ac{CC} with identical channel realizations outperforms \ac{PC} until reaching the maximum detection probability. Then, the superiority of \ac{CC} is also affected by the number of receiving antennas and repetitions. In this case, the largest difference in detection probability for format B1 is found with $n^{(\mathrm{ant})}=8$, while for format B2 it is with $n^{(\mathrm{ant})}=4$.


\begin{figure}[t]
\begin{center}
\includegraphics[width=\columnwidth]{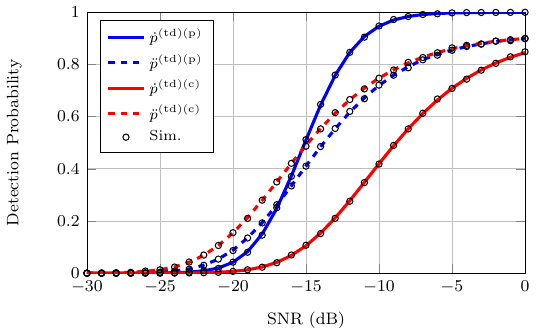}
\caption{Detection probability comparison of combining techniques for format B1 with $n^{(\mathrm{ant})}=1$ and $n^{(\mathrm{inter})}=1$.}
\label{fig:6}
\end{center}
\end{figure}

As noted above, combiners are affected by interference from other roots configured in the same \ac{gNB}. Since $n^{(\mathrm{inter})}$ cannot be known, its average value must be selected based on expected traffic, in order to design the optimal threshold that statistically meets the target $p^{(\mathrm{fa\_des})}$. Fig. \ref{fig:6} represents the case of $n^{(\mathrm{inter})}=1$ for format B1 with $n^{(\mathrm{ant})}=1$. In \ac{PC}, interference simply adds extra power (and a small cross-term), raising the noise floor and reducing $p^{(\mathrm{td})}$. Meanwhile, in \ac{CC}, the interference signal is added to the transmitted signal before squaring. Depending on its phase relative to the transmitted one, it can be combined constructively or destructively. We observe that \ac{CC} in high-\ac{SNR} scenarios has a $p^{(\mathrm{td})}$ that is 0.1 lower than that of the case without interference, while \ac{PC} is only affected in the case of identical channel realizations.

\subsection{Designed algorithm performance with \ac{CFO}}
\label{subsec:DesigAlgoPerfwCFO}

This subsection validates the algorithm designed in the presence of \ac{CFO} with \ac{PC} for formats C0 and 0 with $\epsilon=0.3$. It is also compared to a \textit{conventional \ac{PRACH} detector}, whose threshold is computed based on the estimation of the \ac{PDP} noise floor \cite[Ch. 19]{sesia2011lte}.

\begin{figure}[!ht]
\begin{center}
\includegraphics[width=\columnwidth]{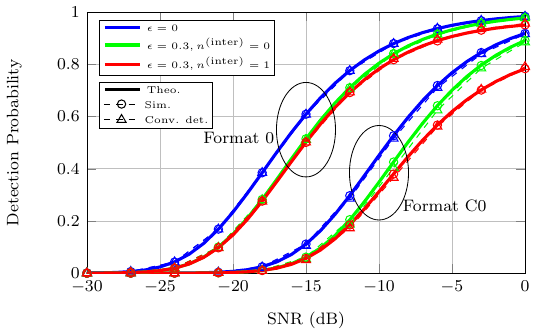}
\caption{Detection probability of formats 0 and C0 for the cases of $\epsilon=0$, $\epsilon=0.3$, and $\epsilon=0.3$ with $n^{(\mathrm{inter})}=1$. Results are compare to a conv. det. \cite[Ch. 19]{sesia2011lte}.}
\label{fig:7}
\end{center}
\end{figure}

A $p^{(\mathrm{td})}$ comparison between the two formats can be seen in Fig. \ref{fig:7} for different scenarios. We observe that format 0 achieves a better $p^{(\mathrm{td})}$ than format C0 since it has a higher $\ell$, collecting more signal power when computing the \ac{PDP}. On the other hand, we can observe how when introducing \ac{CFO}, the performance worsens as the threshold has to be increased to maintain the target $p^{(\mathrm{fa})}$, also in the case of interfering devices. Our detector algorithm and the conventional one (conv. det.) obtain similar performance in terms of $p^{(\mathrm{td})}$, since the latter is capable of detecting any peak crossing a threshold. However, it cannot distinguish between true and false peaks originated from \ac{CFO}, as we will see below.


\begin{figure}[!ht]
\begin{center}
\includegraphics[width=\columnwidth]{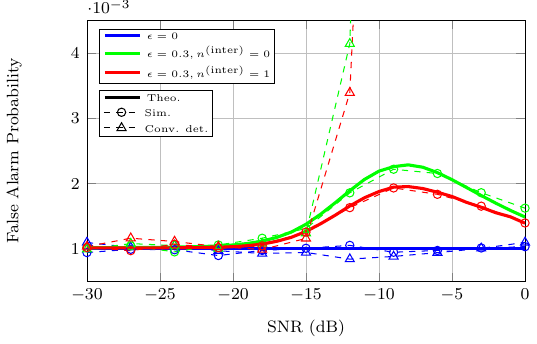}
\caption{False alarm probability of format C0 for the cases of $\epsilon=0$, $\epsilon=0.3$, and $\epsilon=0.3$ with $n^{(\mathrm{inter})}=1$. Results are compare to a conv. det. \cite[Ch. 19]{sesia2011lte}.}
\label{fig:8}
\end{center}
\end{figure}

\begin{figure}[!ht]
\begin{center}
\includegraphics[width=\columnwidth]{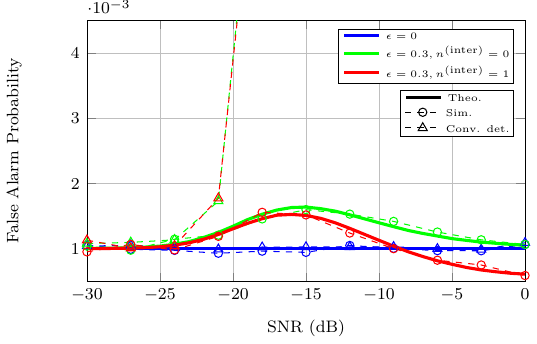}
\caption{False alarm probability of format 0 for the cases of $\epsilon=0$, $\epsilon=0.3$, and $\epsilon=0.3$ with $n^{(\mathrm{inter})}=1$. Results are compare to a conv. det. \cite[Ch. 19]{sesia2011lte}.}
\label{fig:9}
\end{center}
\end{figure}

Figs. \ref{fig:8} and \ref{fig:9} represent the $p^{(\mathrm{fa})}$ of formats C0 and 0, respectively, for different scenarios. In both, the probability remains as desired in the case of no \ac{CFO}, while with \ac{CFO} it increases with the \ac{SNR} up to a certain point and then decreases, making a maximum of $p^{(\mathrm{fa})}$. For the no-interfering case, this effect is due to the fact that as \ac{SNR} increases, the probability of a false peak exceeding the true peal also increases. This effect goes up to a certain \ac{SNR} beyond which the signal is large enough that the effect of noise can be neglected, and thus the probability returns to the desired level. In the case of interfering devices, it is similar, except that it does not return to the desired level but continues to decrease. As we increase the \ac{SNR} the base threshold, $\dot{\mathcal{T}}_{u_{0}}^{(\text{p})}$, starts to become more restrictive than the CFO threshold, $\dot{\mathcal{T}}^{(\mathrm{CFO})}_{u_{0}}$, removing more noise-detected peaks than necessary to maintain the desired probability. This effect can be mitigated, as observed in the subsequent results. 

Note that the nominal frequency offset that each format can tolerate differs greatly. Format 0 with $\epsilon=0.3$ corresponds to a nominal offset of 375 Hz, which corresponds to a user speed of 120 km/h with a carrier frequency of 3.5 GHz. Whereas, for the format C0 it is 4.5~kHz or the same user speed with a carrier frequency of 40.5 GHz.  It is important to note that in both cases the conventional detector cannot maintain the $p^{(\mathrm{fa})}$, since it is not able to distinguish between false and true peaks. The conventional detector increases its $p^{(\mathrm{fa})}$ with \ac{SNR} until all \ac{PRACH} occasions result in false alarms.


\begin{figure}[!ht]
\begin{center}
\includegraphics[width=\columnwidth]{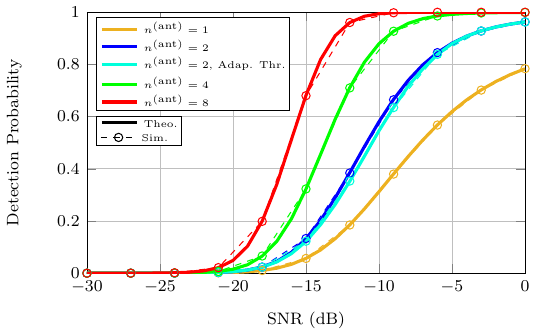}
\caption{Detection probability of format C0 for different numbers of received antennas with $n^{(\mathrm{inter})}=1$. The case of threshold adaptation for 2 antennas is also represented.}
\label{fig:10}
\end{center}
\end{figure}

\begin{figure}[!ht]
\begin{center}
\includegraphics[width=\columnwidth]{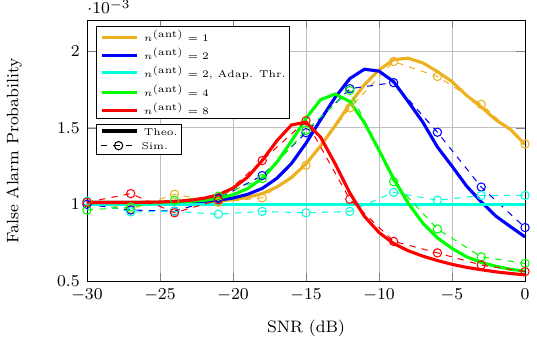}
\caption{False alarm probability of format C0 for different numbers of received antennas with $n^{(\mathrm{inter})}=1$. The case of threshold adaptation for 2 antennas is also represented.}
\label{fig:11}
\end{center}
\end{figure}

Fig. \ref{fig:10} and \ref{fig:11} show the $p^{(\mathrm{td})}$ and $p^{(\mathrm{fa})}$ of format C0 as the number of antennas increases, with $n^{(\mathrm{inter})}=1$, respectively. Every time we double the number of antennas, the same $p^{(\mathrm{td})}$ is achieved with a 3 dB lower \ac{SNR}, exactly matching the expected noise reduction from averaging i.i.d. noise. We also see that the $p^{(\mathrm{fa})}$ due to \ac{CFO} and its maximum are also reduced. On the other hand, the results of using an adapted base threshold computed by compensating for all the effects introduced by the \ac{CFO} and the interference are also plotted for the case of $n^{(\mathrm{ant})}=2$. With this threshold, we can hold $p^{(\mathrm{fa})}$ at its target level without significantly degrading $p^{(\mathrm{td})}$, improving spectral efficiency by reducing the number of attempts required or the reserve of wasted resources.


\section{Conclusions}
\label{sec:Conclusions}

We have developed and validated a comprehensive statistical framework for 5G-\ac{NR} \ac{PRACH} preamble detection and derived closed‑form \ac{PDP} distributions under both hypotheses that account for multiple receive antennas, multiple simultaneous devices, multiple roots, \ac{CFO}, and two combining methods. We identify two key cases: identical and independent channel realizations per repetition, that markedly change \ac{PRACH} performance, and providing threshold expressions dependent on cell traffic to simplify design and analysis. We show that \ac{PC} is more robust under independent repetition channels and interference, while \ac{CC} performs best in low-\ac{SNR} scenarios when repetitions experience identical channels. Finally, we proposed a novel detector with an adaptive threshold that has the same detection capabilities as a conventional detector, while tightly controlling the false alarm probability as desired, and thus its spectral efficiency. The extension of the proposed framework to multipath channels is left for future work, as its exhaustive and comprehensive explanation requires a dedicated study. These results provide both theoretical insight and practical tools for efficient \ac{PRACH} configuration in dense and heterogeneous 5G deployments.


\section{Appendix. Useful distributions} 
\label{sec:Distributions}
\Acf{RV} X is distributed following the known distribution $\text{D}$, $\text{X} \sim \text{D}$, e.g. $\text{X} \sim \mathcal{N}(\mu,\sigma^2)$. The  \ac{PDF} is denoted as $f_{\text{X}} (x) = f_{\text{D}} (x)$, the \ac{CDF} as $F_{\text{X}} (x) = F_{\text{D}} (x)$, the \ac{CCDF} as $\bar{F}_{\text{X}}(x) = 1-F_{\text{X}}(x) = \bar{F}_{\text{D}} (x)$ and the \ac{MGF} is $M_{\text{X}}(t)=M_{\text{D}}(t)={\mathbb{E}}[\exp(t\text{X})]$. For vectors, each element is distributed according to the distribution, and we assume independence between elements unless otherwise stated. \Acp{RV} often depend on other random variables. We express that dependency as $\text{X}(\text{Y})$. For vectors, $\mathbf{x}(\mathbf{y})$, e.g., $\text{X}({\mathbf{h}}) \sim \mathcal{CN}(\mu ({\mathbf{h}}),\sigma^2)$.



\subsection{Scaling and translation}
\label{subsec:scaling}
If $\text{X} \sim \text{D}$, and $\text{Y} = \sigma^{2}  \text{X} + \mu$, with $ \sigma^{2} > 0$, then the \ac{PDF}, \ac{CDF} and \ac{MGF} can be expressed as:

\begin{equation}\label{scaleDistr}
\begin{array}{@{}ll@{}}
f_{\text{Y}} (y)
  &= \frac{1}{{\sigma^{2}}}f_{\text{D}} \left( {\frac{{y - \mu}}{\sigma^{2}}} \right), 
   \quad
F_{\text{Y}} (y)
  = \displaystyle F_{\text{D}} \left( {\frac{{y - \mu}}{\sigma}} \right), \\[1ex]
M_{\text{Y}} (t)
  &= e^{\mu t} M_{\text{D}} \left( {\sigma^{2} t} \right). 
\end{array}
\end{equation}

\subsection{Chi-squared ($\chi^2$) Distribution}
\label{subsec:Chisquared}
If $\text{X} = \sum\limits_{i = 1}^K {\text{N}_i ^2 }$ with $\text{N}_i \sim \mathcal{N}(0,1)$, then $\text{X}$ follows a Chi-squared distribution with $K$ degrees of freedom, $\text{X} \sim \chi_K^2$. The \ac{PDF}, \ac{CCDF} and \ac{MGF} are given by:
\begin{equation}\label{eq:ChiSquaredDistr}
\begin{array}{@{}ll@{}}
f_{\chi_{2\alpha}^2}(x)
  &= \displaystyle\frac{x^{\alpha-1}e^{-x/2}}{2^\alpha\,\Gamma(\alpha)} , 
   \quad
\bar F_{\chi_{2\alpha}^2}(x)
  = \displaystyle\frac{\Gamma\bigl(\alpha,\tfrac{x}{2}\bigr)}{\Gamma(\alpha)}, \\[1ex]
M_{\chi_{2\alpha}^2}(t)
  &= (1-2t)^{-\alpha}, 
   \quad t<\tfrac12,
\end{array}
\end{equation}
with $\Gamma(n,x)$ the upper incomplete gamma function.




\subsection{Non-central $\chi^2$ Distribution ($\chi^{'2}$)}
\label{subsec:NoncentralChiSquared}
If $\Psi  = \sum\limits_{i = 1}^\alpha  {\left| {\dddot Z_i } \right|^2 } $ with $\dddot Z_i \sim \mathcal{CN}(\mu _i ,1)$, then $\Psi$ is distributed following a non-central $\chi^2$ distribution with $2\alpha$ degrees of freedom and $\lambda=\sum\limits_{i = 1}^\alpha  {\left| {\mu _i } \right|^2 }$ as non-centrality parameter. The \ac{PDF}, \ac{CCDF} and the \ac{MGF} of the non-central $\chi^2$ distribution are written as: 
\begin{equation}
\label{eq:pdfnoncentralChi2}    
f_{\chi _{2\alpha }^{'2} } (x;\lambda ) = \frac{1}
{2}e^{ - \frac{{x + \lambda }}
{2}} \left( {\frac{x}
{\lambda }} \right)^{\frac{{\alpha  - 1}}
{2}} I_{\alpha  - 1} \left( {\sqrt {\lambda x} } \right),\quad x > 0
\end{equation}
\begin{equation}
\label{eq:ccdfnoncentralChi2}    
\bar F_{\chi _{2\alpha }^{'2} } (x;\lambda ) = {\text{Q}}_\alpha  \left( {\sqrt \lambda  ,\sqrt x } \right),x > 0
\end{equation}
\begin{equation}
\label{eq:mgfnoncentralChi2}    
{M}_{\chi _{2\alpha }^{'2} } (t;\lambda ) = (1 - 2t)^{ - \alpha } \exp \left( {\frac{{\lambda t}}
{{1 - 2t}}} \right),
\end{equation}
with ${\text{Q}}_M  \left(a,b\right )$ the Marcum Q-function of order M. 

\subsection{Scaled $\chi^{'2}$ Distribution}
\label{subsec:ScaledNoncentralChiSquared}
If  $\Psi  = \sum\limits_{i = 1}^\alpha  {\left| {\text{Z}_i } \right|^2 } $ with $\text{Z}_i \sim \mathcal{CN}(\mu _i ,\sigma^2)$, then $\Psi$ is distributed following a scaled non-central $\chi^2$ distribution with $2\alpha$ degrees of freedom and ${\lambda  = \sum\limits_{i = 1}^\alpha \left| \mu_{i} \right|^2/\sigma^{2}  }$ as non-centrality parameter, that is, $\Psi \sim \sigma^2 \chi _{2\alpha }^{'2} \left( {\sum\limits_{i = 1}^\alpha \left| \mu_{i} \right|^2/\sigma^{2}} \right)$. The \ac{PDF}, \ac{CCDF}, and \ac{MGF} of the scaled noncentral $\chi^2$ distribution can be obtained by applying the scale expressions in Subsection \ref{subsec:scaling} to those in Subsection \ref{subsec:NoncentralChiSquared}.

\subsection{Scaled $\chi^{'2}$ distribution with scaled $\chi^2$ non-centrality parameter}
\label{subsec:ncscaled-scaledchi2}
Let $\text{X} \sim \sigma_x^2 \chi _{2\alpha }^{'2} (\Lambda)$ be a scaled non-central chi-square random variable with $2\alpha$ \ac{DOF} and non-centrality parameter $\Lambda$. Assume that the non-centrality itself is random, $\Lambda \sim \sigma _\lambda ^2 \chi _{2\beta }^2$, where $\Lambda$ is a scaled central chi-square with $2\beta$ \ac{DOF}. 
Then, X follows a generalized $\chi^2$ distribution:
\begin{equation}
\label{eq:dist_ncscaled-scaledchi2}
\text{X} \sim \sigma _x^2 \chi _{2(\alpha  - \beta )}^2 * \sigma _x^2(1  + \sigma _\lambda ^2 )\chi _{2\beta }^2,
\end{equation}
with $*$ is the convolution operator coming from the addition of two random variables.


Consequently, its \ac{CDF} is given by:
\begin{multline}
\label{eq:CDFofX}
    F_{\text{X}} (x) = \\ 
    = \sum_{j=0}^{\infty} \frac{\Gamma(j+\beta)}{j!\Gamma(\beta)} \left( \frac{\sigma _\lambda ^2}{\sigma _\lambda ^2 +1} \right)^{j}
     \left( \frac{1}{\sigma _\lambda ^2 + 1} \right)^{\beta} \frac{\gamma \left( \alpha+j, \frac{x}{2\sigma _x^2} \right)}{\Gamma(\alpha+j)},
\end{multline}
where $\gamma(\cdot,\cdot)$ is the lower incomplete gamma function.

In case that $\alpha=\beta$, the result is reduced to an unique scaled $\chi^2$ distribution, ${\text{X}} \sim \sigma _x^2(1  + \sigma _\lambda ^2 )\chi _{2\alpha }^2$, with corresponding CDF:
\begin{equation}
\label{eq:CDFofChisquare}
    F_{\text{X}} (x) =  \frac{\gamma \left( {\alpha ,\frac{x} {2 \sigma _x^2(1  + \sigma _\lambda ^2 )}} \right)} {{\Gamma (\alpha )}}.
\end{equation}


%
\ifCLASSOPTIONcaptionsoff
  \newpage
\fi
\bibliographystyle{IEEEtran} 
\bibliography{biblio}

@TECHREPORT{38.211,
  author = {3GPP},
  title = {{5G; NR; Physical channels and modulation}},
  institution = {{3rd Generation Partnership Project (3GPP)}},
  year = {2021},
  type = {TS},
  number = {{38.211 v16.07.00}},
  month = {October}
}

@STANDARD{3gpp38141-1,
  title = {{NR; Base Station (BS) conformance testing; (Release 18)}},
  author  = {{3rd Generation Partnership Project (3GPP)}},
  type = {TS},
  number = {38.141-1 V18.10.0},
  month = {July},
  year = {2025}
}

@STANDARD{38.213,
  title = {{5G; NR; Physical layer procedures for control}},
  author = {{3rd Generation Partnership Project (3GPP)}},
  type = {TS},
  number = {38.213 v18.6.0},
  month = {April},
  year = {2025}
}

@STANDARD{3gpp38211,
  title = {{NR; Physical channels and modulation; (Release 18)}},
  author = {{3rd Generation Partnership Project (3GPP)}},
  type = {TS},
  number = {38.211 V18.2.0},
  month = {September},
  year = {2023}
}

@book{sesia2011lte,
  title={LTE-the UMTS long term evolution: from theory to practice},
  author={Sesia, Stefania and Toufik, Issam and Baker, Matthew},
  year={2011},
  publisher={John Wiley \& Sons}
}

@article{chakrapani2020design,
  title={On the design details of {SS/PBCH}, signal generation and {PRACH} in {5G-NR}},
  author={Chakrapani, Arvind},
  journal={IEEE Access},
  volume={8},
  pages={136617--136637},
  year={2020},
  publisher={IEEE}
}

@article{pitaval2020overcoming,
  title={Overcoming {5G} {PRACH} capacity shortfall: Supersets of {Zadoff--Chu} sequences with low-correlation zone},
  author={Pitaval, Renaud-Alexandre and Popovi{\'c}, Branislav M and Wang, Peng and Berggren, Fredrik},
  journal={IEEE Transactions on Communications},
  volume={68},
  number={9},
  pages={5673--5688},
  year={2020},
  publisher={IEEE}
}

@article{liang2017non,
  title={Non-orthogonal random access for {5G} networks},
  author={Liang, Yanan and Li, Xu and Zhang, Jiayi and Ding, Zhiguo},
  journal={IEEE Transactions on Wireless Communications},
  volume={16},
  number={7},
  pages={4817--4831},
  year={2017},
  publisher={IEEE}
}

@article{ding2019analysis,
  title={Analysis of non-orthogonal sequences for grant-free {RA} with massive {MIMO}},
  author={Ding, Jie and Qu, Daiming and Choi, Jinho},
  journal={IEEE Transactions on Communications},
  volume={68},
  number={1},
  pages={150--160},
  year={2019},
  publisher={IEEE}
}

@article{tao2018improved,
  title={Improved {Zadoff-Chu} sequence detection in the presence of unknown multipath and carrier frequency offset},
  author={Tao, Jun and Yang, Le},
  journal={IEEE Communications Letters},
  volume={22},
  number={5},
  pages={922--925},
  year={2018},
  publisher={IEEE}
}

@article{hua2013analysis,
  title={Analysis of the frequency offset effect on random access signals},
  author={Hua, Min and Wang, Mao and Yang, Wenjie and You, Xiaohu and Shu, Feng and Wang, Jianxin and Sheng, Weixing and Chen, Qian},
  journal={IEEE Transactions on communications},
  volume={61},
  number={11},
  pages={4728--4740},
  year={2013},
  publisher={IEEE}
}

@article{wang2015multiuser,
  title={A multiuser detection algorithm for random access procedure with the presence of carrier frequency offsets in {LTE} systems},
  author={Wang, Qiwei and Ren, Guangliang and Wu, Jueying},
  journal={IEEE Transactions on Communications},
  volume={63},
  number={9},
  pages={3299--3312},
  year={2015},
  publisher={IEEE}
}

@inproceedings{yang2013enhanced,
  title={Enhanced preamble detection for {PRACH} in {LTE}},
  author={Yang, Xiaobin and Fapojuwo, Abraham O},
  booktitle={2013 IEEE Wireless Communications and Networking Conference (WCNC)},
  pages={3306--3311},
  year={2013},
  organization={IEEE}
}

@article{kim2017enhanced,
  title={An enhanced {PRACH} preamble detector for cellular {IoT} communications},
  author={Kim, Taehoon and Bang, Inkyu and Sung, Dan Keun},
  journal={IEEE Communications Letters},
  volume={21},
  number={12},
  pages={2678--2681},
  year={2017},
  publisher={IEEE}
}

@article{chakrapani2019nb,
  title={{NB-IoT} uplink receiver design and performance study},
  author={Chakrapani, Arvind},
  journal={IEEE Internet of Things Journal},
  volume={7},
  number={3},
  pages={2469--2482},
  year={2019},
  publisher={IEEE}
}

@inproceedings{kim2016transmit,
  title={Transmit power optimization for prioritized random access in {OFDMA} based systems},
  author={Kim, Taehoon and Ko, Kab Seok and Sung, Dan Keun},
  booktitle={2016 IEEE International Conference on Communications (ICC)},
  pages={1--6},
  year={2016},
  organization={IEEE}
}

@article{zhu2023timing,
  title={Timing advance estimation in low earth orbit satellite networks},
  author={Zhu, Jianfeng and Sun, Yaohua and Peng, Mugen},
  journal={IEEE Transactions on Vehicular Technology},
  volume={73},
  number={3},
  pages={4366--4382},
  year={2023},
  publisher={IEEE}
}

@book{enescu20205g,
  title={5G New Radio: A beam-based air interface},
  author={Enescu, Mihai},
  year={2020},
  publisher={John Wiley \& Sons}
}

@book{berndt1998gauss,
  title={Gauss and Jacobi sums},
  author={Berndt, Bruce C and Evans, Ronald J and Williams, Kenneth S},
  year={1998},
  publisher={Wiley}
}

@article{patnaik1949non,
  title={The non-central $\chi$ 2-and F-distribution and their applications},
  author={Patnaik, PB},
  journal={Biometrika},
  volume={36},
  number={1/2},
  pages={202--232},
  year={1949},
  publisher={JSTOR}
}

@inproceedings{ota2023nr,
  title={{NR} {PRACH} Detection Probability in the Presence of {CFO} in Multi-access Environments},
  author={Ota, Yoshinobu and Chiba, Takamichi and Sawahashi, Mamoru and Suyama, Satoshi},
  booktitle={2023 VTS Asia Pacific Wireless Communications Symposium (APWCS)},
  pages={1--5},
  year={2023},
  organization={IEEE}
}

\end{document}